\documentclass[manuscript, screen]{acmart}

\usepackage{xcolor} 
\usepackage{lscape} 
\usepackage{multirow}
\usepackage{enumitem}
\usepackage{wrapfig}
\usepackage{rotating}
\usepackage{epstopdf}

\AtBeginDocument{%
  \providecommand\BibTeX{{%
    \normalfont B\kern-0.5em{\scshape i\kern-0.25em b}\kern-0.8em\TeX}}}

\setcopyright{acmcopyright}
\copyrightyear{2018}
\acmYear{2023}
\acmDOI{XXXXXXX.XXXXXXX}

\acmJournal{CSUR}
\acmVolume{37}
\acmNumber{4}
\acmArticle{111}
\acmMonth{8}

\begin{document}

\title{SoK: Demystifying Privacy Enhancing Technologies Through the Lens of Software Developers}

\author{Maisha Boteju}
\affiliation{%
  \institution{The University of Auckland}
  \city{Auckland}
  \country{New Zealand}}
\email{mbot450@aucklanduni.ac.nz}

\author{Thilina Ranbaduge}
\affiliation{%
  \institution{Data61, CSIRO}
  \city{Canberra}
  \country{Australia}
}
\email{thilina.ranbaduge@anu.edu.au}

\author{Dinusha Vatsalan}
\affiliation{%
 \institution{Macquarie University}
 \city{Sydney}
 \country{Australia}}
 \email{dinusha.vatsalan@mq.edu.au}

\author{Nalin Asanka Gamagedara Arachchilage}
\affiliation{%
  \institution{The University of Auckland}
  \city{Auckland}
  \country{New Zealand}}
 \email{nalin.arachchilage@auckland.ac.nz }
        
\renewcommand{\shortauthors}{Boteju et al.}

\begin{abstract}
In the absence of data protection measures, software applications lead to privacy breaches, posing threats to end-users and software organisations. Privacy Enhancing Technologies (PETs) are technical measures that protect personal data, thus minimising such privacy breaches. However, for software applications to deliver data protection using PETs, software developers should actively and correctly incorporate PETs into the software they develop. Therefore, to uncover ways to encourage and support developers to embed PETs into software, this Systematic Literature Review (SLR) analyses 39 empirical studies on developers' privacy practices. It reports the usage of six PETs in software application scenarios. Then, it discusses challenges developers face when integrating PETs into software, ranging from intrinsic challenges, such as the unawareness of PETs, to extrinsic challenges, such as the increased development cost. Next, the SLR presents the existing solutions to address these challenges, along with the limitations of the solutions. Further, it outlines future research avenues to better understand PETs from a developer perspective and minimise the challenges developers face when incorporating PETs into software.

\end{abstract}


\maketitle

\section{Introduction}\label{introduction}
Software applications pose a significant threat to the privacy of their end-users \cite{moneymakes, agile, platformprivacies}. They routinely and excessively collect personal data, such as personally identifiable information (e.g., names, locations) and sensitive data (e.g., healthcare data), in exchange for their services \cite{goingbeyond, aiethicspractice}. For instance, in 2018, the New York Times found that some software track the location of 200 million mobile users in the United States, collecting accurate location data over 14,000 times daily \cite{nytimes}. Therefore, if software applications do not provide means to protect these personal data, end-users could lose them to unknown parties who might sometimes be malicious, causing end-users financial losses, reputation damage and emotional distress \cite{devcannotembed,willtheyuse}. For example, the Ashley Madison website, a platform facilitating extramarital affairs, fell victim to a hacking incident that exposed the names, contact details, locations and financial transactions of 30 million of its users \cite{ashley}. As a result, the reputation of some celebrities and government officials was damaged, more divorce cases were triggered, and some users even committed suicide \cite{ashley}. This incident exemplifies how the absence of data protection within software applications can profoundly affect its end-users.

Therefore, developing software applications with adequate data protection measures is essential to protect end-users from such harmful consequences. With scrutinised privacy regulations (e.g., GDPR \cite{gdprpseudo}) in place and increased public awareness about privacy, software organisations are pressured to safeguard end-user data more than ever \cite{agile, darn, gamedevs, senarath2021unheard}. To meet this demand, integrating Privacy Enhancing Technologies (PETs) into software emerges as a crucial strategy to achieve data protection \cite{medco, tippers}. PETs are a collection of technical solutions which allow software applications to extract insights from personal data while upholding data protection principles such as minimising the use of personal data, i.e., data minimisation \cite{petsgovernanddesign}. For instance, federated learning is a PET that allows training a machine learning model using decentralised data (data splitting), ensuring those data are kept private at their original locations \cite{towardsfl, aiethicspractice}. PETs achieve data protection through different approaches, including transforming data to de-identify individuals, hiding or shielding data, and splitting or controlling access to data \cite{PETsICO}. Using these approaches, PETs can achieve data protection by design and default in software, proactively preserving the privacy of the data owners \cite{12, PETsICO}.

However, software developers rarely consider using PETs to ensure data protection in software \cite{devcannotembed}. They usually provide privacy policies and settings or limit data collection \cite{darn, mindset}. However, these techniques do not guarantee that personal data are protected once the data are disclosed to software. Moreover, developers struggle to make decisions when embedding PETs into software \cite{dmapproach}. They are unsure how to select suitable PETs for their applications or when and how to use them \cite{devcannotembed,dontlook,petsgovernanddesign}. In addition, they constantly require external verification for their decisions, showing a lack of confidence in working with PETs \cite{devcannotembed}. As a result of this behaviour, developers fail to create software applications with robust data protection guarantees, compromising end-user privacy and holding organisations vulnerable to regulatory pressure \cite{devcannotembed}. Therefore, it is important to change developers' behaviour in accepting PETs and correctly integrating them into software.

To identify possible approaches to support this behavioural change, first, it is required to understand how developers are currently using PETs, what makes the integration of PETs challenging for them, and how these challenges can be addressed. However, few empirical studies are dedicated to exploring the developer perspective of utilising PETs \cite{dontlook, petsgovernanddesign}. In many cases, PETs are only a brief part of empirical studies that explore developers' privacy practices \cite{devcannotembed, willtheyuse, insidetheorg, aiethics, aiethicspractice}. Therefore, synthesising the knowledge in those empirical studies, this Systematic Literature Review (SLR) provide a holistic understanding of how to support developers in incorporating PETs into software. The following research questions guide the SLR.

\begin{enumerate}
  \item[RQ1 :] What existing PETs have developers integrated into the software to achieve data protection, and how have these PETs been integrated?
  \item[RQ2 :] What are the challenges developers face when integrating PETs into the software they develop?
  \item[RQ3 :] What solutions does the literature propose to tackle the challenges identified in RQ2, and what are the limitations associated with these solutions?
\end{enumerate}

The remainder of the article is organised as follows. Section \ref{preliminaries} briefly describes the PETs discussed throughout the SLR. Next, Section \ref{related} discusses our contribution to the body of knowledge. In Section \ref{methodology}, we outline the methodology of the SLR followed by the findings of the SLR in Section \ref{results}. Then Section \ref{discussion} presents the implications of the results, along with future research areas and limitations of the SLR. Finally, Section \ref{conclusion} presents the concluding remarks of the SLR.

\section{Preliminaries} \label{preliminaries}
This section briefly describes the types of PETs discussed throughout our SLR to guide the readers. We also outline the data protection capabilities of the discussed PETs by explaining how they can be integrated into a hypothetical software application scenario.\\

\noindent \textbf{Software application scenario}: A software company is developing a web-based EHR (Electronic Health Record) system for three hospitals (A, B, C) to manage patient data. The doctors and nurses can enter patient's data into the system, including name, age, zip code, race, gender, contact number, symptoms, and diagnosis. Since healthcare data contain sensitive personal data, the collected data are encrypted at storage. The EHR system is hosted on Amazon Web Services (AWS), utilising its cloud infrastructure, storage, and services for enhanced performance and scalability. The system incorporates machine learning for diagnosis prediction. Further, it facilitates secure data sharing with external research institutions. In addition, insurance companies can register with the EHR system to receive patient health data for claims approval. Finally, the EHR system enables the three hospitals to collaboratively analyse patient data handled by their respective EHR systems to obtain rich insights. However, they do not want to share their data during this collaborative process.

\subsubsection{Psuedonymisation} Psuedonymisation is a de-identification technique where an individual's personally identifiable information (PII) is replaced by pseudonyms (artificial identifiers) \cite{goingbeyond}. According to the General Data Protection Regulation (GDPR), organisations utilising personal data should employ pseudonymisation to achieve secure data storage \cite{gdprpseudo}. In the given application scenario, pseudonymisation can be used in database tables, where the actual name of each patient is replaced by a pseudonym patient\# (\# is a number different for each patient). In this way, linking health data to the respective patients is minimised, thus improving data protection in the application \cite{PETsICO}.

\subsubsection{K-Anonymity} K-anonymity is one of the popular forms of syntactic anonymisation, in which anonymisation is achieved by changing the appearance of data \cite{anonsweeny}. A tabular dataset is said to be k-anonymised if a set of quasi-identifiers \footnote{Quasi-identifiers are data attributes, excluding PII, that in combination can uniquely identify a person \cite{anonsweeny}. The quasi-identifiers of the given application scenario are the age, zip code, race, and gender of a patient.} are repeating at least in k different rows \cite{anonsweeny}. K- anonymity can be achieved either through \textbf{generalisation} (replacing an individual data value with a more general value) or \textbf{suppression} (removing data values in a dataset to achieve anonymity) \cite{lessonslearned, anonsweeny}. In the given application, the "race" data can be removed from the dataset if it is not necessary for the functionality of the application (suppression), the last three digits of the zip code can be replaced with the `*' symbol (generalisation) and the age can be stored as a numerical range, for example, 25 can be stored as 20-30 age range (generalisation). These steps must be completed so that values for age, zip code, race, and gender in the dataset repeat at least in k rows \cite{anonsweeny}. In this manner, the specificity of the patient data is minimised, thus minimising their identifiability \cite{PETsICO}.

\subsubsection{Onion Routing} Onion routing is used to camouflage the communication that occurs in public networks \cite{onionrouting}. This technique routes network users' encrypted messages through a series of intermediate servers before reaching their destination \cite{onionrouting}. In this way, the data communicated and the information about who is communicating with whom are hidden \cite{onionrouting}. In the given example, onion routing can be used to anonymise the communication between the end-users (doctors, nurses, and insurance companies) and the web-based EHR system.

\subsubsection{Homomorphic Encryption} Homomorphic encryption (HE) is an encryption technique which supports computation over encrypted data without decrypting them \cite{medco, videosharing}. In the given application scenario, suppose a doctor in hospital A must know the number of COVID-19 patients who visited the hospital within a specific time period. If the EHR system has integrated HE, it can perform addition on the encrypted data to return the total COVID-19 patients to the doctor. The result will be decrypted on the client side of the system. This way, personal data are hidden even during the computations, thus enhancing the data protection of the system \cite{PETsICO}.

\subsubsection{Federated Learning} Federated learning (FL) can perform machine learning using remotely existing data without bringing them to one location \cite{towardsfl}. In this technique, a central server contains a global machine-learning model and copies of this model is sent to every remote node that contains the training data \cite{fldefinition}. These copies (local models) are then trained locally at the remote nodes using the data stored in them \cite{fldefinition}. The updates done to the local models are communicated back to the global model, where it aggregates the updates from all remote nodes \cite{fldefinition}. This process is repeated throughout the training process. In the given application, the three hospitals can utilise FL to train a patient similarity detection model to determine clinical trial participants \cite{patientsimi}. In this manner, the data collected separately by the hospitals is kept private at their original sites while valuable insights are extracted from them \cite{PETsICO}.

\subsubsection{Differential Privacy} Differential privacy (DP) is a cryptographic technique where random noise is added to the data query results of a dataset \cite{dontlook,epsilon}. The amount of noise added must make minimal changes to the query results when an individual's data is added or removed from the dataset \cite{epsilon}. In this manner, it is challenging to identify the respective owners of the data included in the dataset \cite{PETsICO}. DP is a semantic anonymisation technique, where anonymisation is achieved by changing the meaning of data \cite{goingbeyond}. However, the added noise decreases the accuracy of the data. Therefore, DP would be much more suitable for systems that analyse aggregated statistical results in data (e.g., census data analysis) rather than for systems which depends on accurate individual-level analysis (e.g., tax calculation) \cite{PETsICO,lessonslearned}. In the example application, suppose the doctors are required to identify the number of diabetic patients based on different age ranges (e.g., 10-15, 16-20). The system can utilise DP to query this, assuring data protection. In this way, even if the results are less accurate due to the added noise, the doctors can understand the relationship between people's age and the likelihood of developing diabetes through the aggregated results.

\subsubsection{Synthetic Data} Synthetic data is artificially generated data that mimics the statistical properties of real data. Different techniques to generate synthetic data include machine learning, deep learning, and agent-based modelling \cite{syntheticdatagen}. In the given application scenario, the EHR system can provide a feature to generate a synthetic dataset replacing the real patient data so that the data can be securely shared with external research institutes without compromising the sensitive personal data of the patients.

\subsubsection{Secure Multi-Party Computation} Multiple parties can jointly perform a computation over a function using their private inputs through secure multi-party computation (SMPC) \cite{petsgovernanddesign}. In SMPC, every party contributing to the computation only knows their own input and the output generated through the computation \cite{petsgovernanddesign}. Suppose in the given application scenario, hospitals A, B, and C are required to jointly identify people who inject drugs (PWID) that frequently visit their emergency wards. For this, the EHR system can use SMPC to use emergency ward patient lists in each hospital and identify the common PWID patients who visited all three hospitals without sharing the patient lists among the hospitals.

\subsubsection{Zero-Knowledge Proof} Zero-knowledge proof (ZKP) is a cryptographic technique which enables one party (prover) to convince another party (verifier) about the truthfulness of a statement without exposing additional information bound to the statement \cite{controltracing}. Suppose in the given example, an insurance company must know whether a patient's blood sugar level is higher than a certain amount to approve the patient's insurance claims. Using ZKP, the system only proves to the insurance company whether the patient's blood sugar level is above or below the expected level without sharing additional personal details (e.g., exact blood sugar level or the different tests or medications taken). In this instance, ZKP achieve data protection by minimising the patient's data used for a particular process \cite{PETsICO}.

\subsubsection{Trusted Execution Environment} Trusted execution environment (TEE) is an isolated and tamper-resistant secure area in a computing device's processor \cite{tee}. TEE ensure the integrity and confidentiality of codes and data loaded into it, even from privileged sources such as the operating system \cite{tee,PETsICO}. In the given example scenario, the AWS nitro enclave \footnote{\url{https://aws.amazon.com/ec2/nitro/nitro-enclaves/}} feature can be used to create such isolated computing environments. Then, sensitive processes such as diagnosis prediction can be conducted inside that secure environment, ensuring that sensitive data used for that process and the generated predictions are handled securely.

\section{Related Work} \label{related}
PETs empower software to extract insights from end-user data while minimising the risks of privacy violations \cite{petsgovernanddesign}. Due to this data protection capability of PETs, researchers have continuously explored ways to improve the existing PETs or introduce new and more robust PETs \cite{21,39, tippers,fate,exdra, controltracing}. Thus, PETs are becoming an active research area in the privacy domain, with various secondary studies\footnote{Secondary studies analyse existing data relating to a specific research question to synthesise knowledge and answer that research question. These studies depend on the data collected by primary studies, which are the studies that collect new data to address a specific research question \cite{kitchenham2007}} offering distinct perspectives on them.

Existing secondary studies on PETs have heavily focused on analysing the privacy protection capabilities of PETs along different application domains \cite{3,17,30,20,33}. These studies primarily concentrated on domains such as healthcare \cite{3, 9, 11, 15, 23, 24, 28}, transportation \cite{17, 29, 34}, smart infrastructures \cite{18, 10}, and communication \cite{30}. Further, there exist some studies that discuss more specific application scenarios, such as federated learning for image processing \cite{19}, differential privacy for location-based services \cite{20}, homomorphic encryption to outsource computation in deep learning \cite{25}, searchable encryption for cloud computing \cite{16}, and blockchain-powered homomorphic encryption to analyse biometric data \cite{33}.

On the other hand, some secondary studies attempted to reveal the technical limitations \cite{11} and challenges of PETs \cite{21,27,25,32,38}. A study by Dankar et al. \cite{11} discussed limitations of differential privacy, such as the theoretical nature of algorithm variables and the lack of applicability due to the reduction in data utility \cite{11}. In addition, the technical challenges encountered when deploying PETs were heavily discussed, focusing on federated learning \cite{21, 38, 32}. The studies pointed out the architectural level challenges encountered due to the decentralised design of federated learning, such as communication difficulties between remote devices and the global machine learning model \cite{21, 38, 32}, scalability issues \cite{21, 32} and negative effect on the global model performance due to the heterogeneity in client data \cite{21,38}.

In addition, several existing works attempted to synthesise the foundational knowledge of PETs \cite{1,22,21}. The theoretical presentation of different homomorphic encryption algorithms was discussed in \cite{1, 22}. Further, Lo et al. \cite{21} conducted an in-depth review using 231 federated learning-related studies to understand the background, requirement analysis, implementation, evaluation, and architecture design-based knowledge needed to develop federated learning systems. As another way of structuring and organising PETs-related knowledge, several secondary studies attempted to classify the knowledge of PETs. Yin et al. \cite{3} classified the privacy risks of federated learning under five aspects (who, what, when, where, and why) and provided an additional categorisation of existing federated learning schemes based on the mechanisms they employ, namely encryption-based, perturbation-based, anonymisation-based, and hybrid. A similar mechanism-based taxonomy is also presented in \cite{36}, where it focuses on the de-identification mechanisms in the context of microdata.

Overall, PETs-related secondary studies spanned various facets, including presenting the potential applications under different domains, technical limitations and challenges, and knowledge classifications. While these synthesising directions are valuable, having a holistic view of the developer perspective regarding PETs is also essential. Since developers are leading roles in software development, the adoption of PETs in the software industry lies in developers' capability to incorporate PETs into software \cite{willtheyuse}. Therefore, contrasting to the existing secondary studies, our SLR attempts to understand how developers are currently integrating PETs into software, what challenges they face during the integration of PETs, and what solutions the literature provides to remedy the identified challenges and the limitations of those solutions. The insights of our work are valuable to identify directions to encourage and support developers to create PETs-embedded software, thus fostering a privacy-preserving digital environment.

\section{Methodology}  \label{methodology}
The methodology of this SLR followed the guidelines proposed by Kitchenham and Charters, which offer a scientific and reproducible approach to conducting SLRs in the
Software Engineering domain \cite{kitchenham2007}. According to these guidelines, we organised our SLR along three stages: \emph{planning, conducting} and \emph{reporting}. This section describes the planning and conducting stages in detail, while the reporting stage is presented in Section \ref{results}.

\subsection{Planning}
In the planning stage, we established a protocol defining the components required to conduct this SLR. These components include the research questions required to guide the SLR, the search terms to identify the relevant publications, the data sources to search for relevant publications, and the selection criteria to determine the publications to be included in the SLR. Defining these components before conducting the SLR was essential to reduce the potential researcher bias during the conducting stage \cite{kitchenham2007}.

\subsubsection{Formulating the Research Questions} \label{researchquestions}
Well-formulated research questions are essential to reflect the scope of an SLR accurately \cite{kitchenham2007}. Therefore, we used the PICOC (Population, Intervention, Comparison, Outcome, Context) framework, which introduces five elements to define focused research questions \cite{slrsocial, kitchenham2007}. These elements include the \emph{population} that we are interested in (software developers in our case), \emph{intervention} examined under the SLR (PETs), \emph{comparison} done against the intervention, \emph{outcomes} that we are interested in (data protection), and the \emph{context} considered (which is software development in our case). It is worth noting that the element "comparison" was excluded in this SLR since it does not intend to compare PETs against any other intervention(s). We used these PICOC elements as a guide to identify the keywords required to structure the research questions mentioned in Section \ref{introduction}. Table \ref{picoc} depicts the PICOC elements, their descriptions and the respective keywords identified using those elements.

\begin{table*}
\caption{The PICOC elements, their descriptions \cite{kitchenham2007, slrsocial}, and the keywords derived from the PICOC elements to structure the research questions and define the search strings. N/A = Not Applicable}
\label{picoc}
\small
\begin{tabular}{lp{8.5cm}l}
\toprule PICOC element & Description                                                                                            & Keyword                        \\ \midrule
 Population  &  What is the Software Engineer (SE) role, category of SEs, or the application area the SLR is interested in? & software developer                                                                 \\
              &                                                                                                                                           \\
 Intervention  &  What software engineering method, technology, tool or process is the SLR interested in?                          & Privacy Enhancing Technologies                                            \\
              &                                                                                                                                           \\
 Comparison      &  What software engineering method, technology, tool or process with which
the intervention is being compared?                                              & N/A                                                                 \\\\
                 
 Outcomes      &  What are the important results for practitioners?                                              & data protection                                                                   \\
              &                                                                                                                                           \\
 Context       &  What is the context within which the intervention is delivered?                                                 & software development           \\ \bottomrule
\end{tabular}
\end{table*}

\subsubsection{Formulating the Search Strings} \label{searchstrings}
After the research questions were formulated, we determined the search strings needed to identify publications relevant to the research questions. We first used the keywords outlined in Table \ref{picoc} as the primary search strings. Then, we considered the spelling variations (e.g. anonymisation and anonymization), synonyms (e.g. developer and programmer), abbreviations (e.g. PETs), and closely connected terms (data protection and privacy) of the primary search strings to identify additional search strings. Expanding the collection of search strings ensured that relevant publications were not overlooked \cite{kitchenham2007}. Table \ref{stringsset} in Appendix A contains the entire collection of search strings defined in this SLR.

\subsubsection{Selecting the Data Sources} \label{section:datasource}
As the next step, we selected scholarly data sources from which we could identify relevant publications for the SLR. We considered three types of data sources: \emph{digital databases, conference proceedings}, and \emph{journal proceedings}. We considered the six most cited digital databases for our SLR: \emph{IEEE Xplore \footnote{\url{https://ieeexplore.ieee.org/}}, ACM Digital Library\footnote{\url{https://dl.acm.org/}}, Springer Link \footnote{\url{https://link.springer.com/}}, Scopus\footnote{\url{https://www.scopus.com/}}, Sage\footnote{\url{https://journals.sagepub.com/}}, and Google Scholar\footnote{\url{https://scholar.google.com/}}}. For journal and conference proceedings, we selected the top-tier venues that publishes work in privacy and security, software engineering (SE), human-computer interaction (HCI), and information systems (IS). These research tracks were considered as the research questions of this study involved humans (software developers), software applications, and data protection. We considered 13 conferences and 11 journals to search publications, which are detailed in Table \ref{proceedings}.

\begin{table*}
\footnotesize
\caption{Selected conference and journal proceedings which are categorised according to the respective research track.}
\label{proceedings}
\begin{tabular}{p{2.5cm}p{6.25cm}p{6.25cm}}
\toprule
Track &  Conferences   & Journals    \\                \midrule
\texttt Privacy and Security & \begin{itemize}[left=0pt,topsep=0pt]
\item ACM Computer and Communications Security (CCS) 
\item USENIX Security Symposium (USENIX-Security)   
\item IEEE Symposium on Security and Privacy (S\&P)
\item Network and Distributed System Security (NDSS) 
\item Symposium On Usable Privacy and Security (SOUPS) 
\item Proceedings on Privacy Enhancing Technologies (PoPETs)
\end{itemize}\nointerlineskip
& \begin{itemize}[left=0pt,topsep=0pt]
 \item IEEE Transactions on Information Forensics and Security (TIFS) 
 \item Computers and Security (Comput Secur) 
 \item Security and Communication Networks (Secur. Commun. Netw.) 
 \item ACM Transactions on Privacy and Security (TOPS) 
 \item IEEE Transactions on Dependable and Secure Computing (IEEE Trans. Dependable Secure Comput.) 
 \end{itemize}\nointerlineskip\\
\hline
\texttt Software Engineering (SE) &  \begin{itemize}[left=0pt,topsep=0pt]
 \item International Conference on Software Engineering (ICSE) 
 \item ACM Joint European Software Engineering Conference and Symposium on the Foundations of Software Engineering (ESEC/FSE) 
 \item International Conference on Automated Software Engineering (ASE) 

\end{itemize}\nointerlineskip
& 
 \begin{itemize}[left=0pt,topsep=0pt]  
 \item IEEE Transactions on Software Engineering (TSE) 
\end{itemize}\nointerlineskip 
\\
\hline
\texttt Human Computer Interaction (HCI) & 
 \begin{itemize}[left=0pt,topsep=0pt]   
 \item ACM Conference on Human Factors in Computing Systems (CHI) 
 \item The Web Conference (WWW) 
\end{itemize}\nointerlineskip  
&
 \begin{itemize}[left=0pt,topsep=0pt]    
 \item IEEE Transactions on Mobile Computing (IEEE Trans. Mob. Comput.) 
 \item IEEE Transactions on Human-Machine Systems (IEEE Trans. Hum.-Mach. Syst.) 
  \end{itemize}\nointerlineskip   
\\
\hline
\texttt Information Systems (IS)
&
 \begin{itemize}[left=0pt,topsep=0pt]     
 \item International Conference on Management of Data (SIGMOD) 
 \item Conference on Information and Knowledge Management (CIKM) 
  \end{itemize}\nointerlineskip    
&
 \begin{itemize}[left=0pt,topsep=0pt]      
 \item IEEE Transactions on Industrial Informatics (IEEE Trans. Ind. Inform.) 
 \item Information Sciences (Inf. Sci.) 
 \item Information Fusion (Inf. Fusion) 
  \end{itemize}\nointerlineskip      
\\
\bottomrule
\end{tabular}
\end{table*}

\subsubsection{Study Selection Criteria}\label{selectioncriteria}
In this stage, we defined selection criteria to determine the publications to be included or excluded from the SLR \cite{kitchenham2007}. They were defined based on the research questions, study quality (e.g., peer-reviewed articles), study type (e.g., research articles, review articles, or white papers) and metadata (e.g., language, publication type, and published year) \cite{litreviewinternettopaper}. Table \ref{selectionTable} presents the inclusion criteria (IC) and exclusion criteria (EC) defined for this SLR.

\begin{table*}
\caption{The selection criteria (inclusion and exclusion criteria) used to select the most relevant publications for this SLR. The publications that satisfied all inclusion criteria and none of the exclusion criteria were included in the SLR.}
\label{selectionTable}
\label{selectioncriteria - table}
\begin{tabular}{p{15cm}}

\toprule
Inclusion Criteria                                                                                                                                                                                           \\ \midrule

\textbf{IC1}: The publication answers at least one research question.                                                                                                                                                            \\

\textbf{IC2}: The publication is published in a peer-reviewed venue: \emph{such studies have a certain validity in their findings \cite{seslrquality}. Hence, including them increases the quality of the SLR.}                                      \\

\textbf{IC3}: The publication is the most recent version, if it has previously published versions.                                                                                                                       \\

\textbf{IC4}: The publication is published after 2016: \emph{to make sure that the effect of regulations is reflected, we included studies published from 2016, which is the year that the EU’s GDPR was entered into force \cite{force}.} \\

\textbf{IC5}: The publication presents empirical studies conducted involving software developers.                                                                                                                               \\

\textbf{IC6}: The publication is written in English.                                                                                                   \\ \\\hline
Exclusion Criteria                                                                                            \\ \hline

\textbf{EC1}: The publication is a poster, short paper, ongoing study, concept paper, opinion paper, technical paper, white paper, tutorial, abstract, report, news article, book chapter, secondary study, or tertiary study. \\
\bottomrule
\end{tabular}
\end{table*}

\subsection{Conducting the SLR}
Once we finalised the SLR protocol, we searched for the relevant publications in the selected data sources. Then, the identified publications were evaluated against the decided selection criteria to filter and select the most relevant ones. The key information was then extracted from these selected studies to get familiarised with the findings of the selected publications. Finally, we conducted a thorough qualitative analysis of the studies to answer the defined research questions (R1, R2, and R3). These processes are explained in the following sections.

\subsubsection{Search and Selection Strategy} \label{selecting}

\begin{figure}[hbtp]
    \centering
    \includegraphics[width=\textwidth]{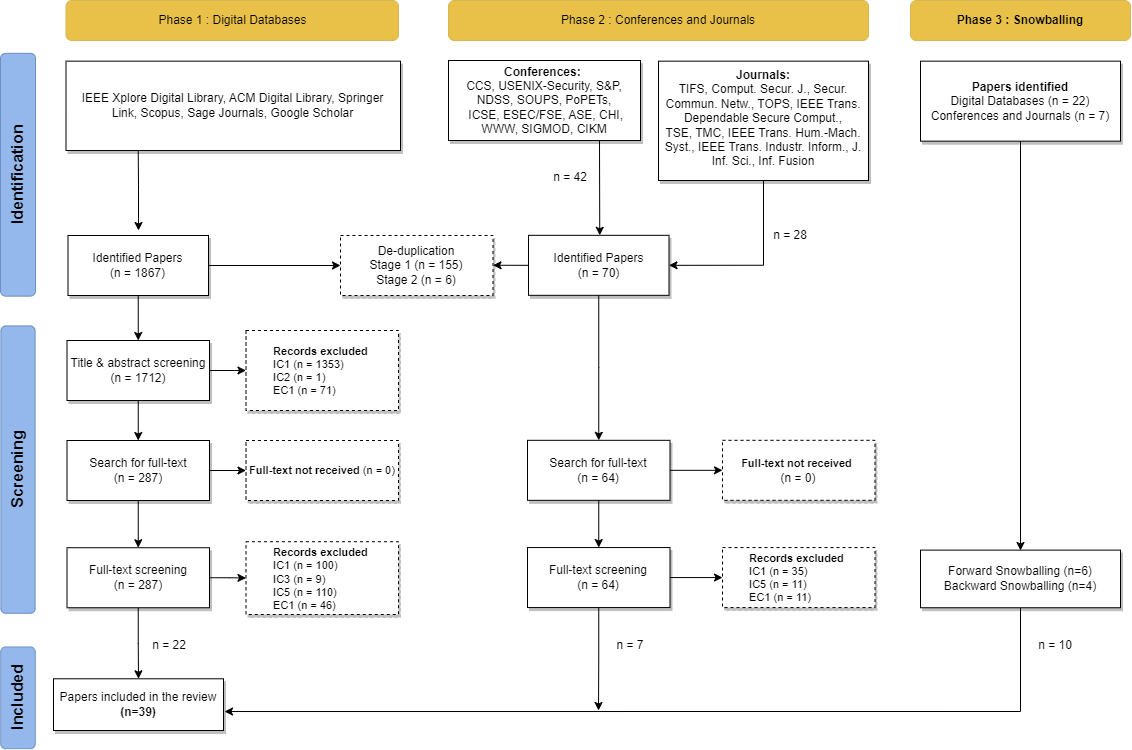}
    \caption{The search and selection strategy of the SLR. The separate phases indicate the three search processes used to identify relevant publications. A total of \textbf{39} primary articles were selected at the end of this process. n = number of papers, IC\# = Inclusion Criteria, EC\# = Exclusion Criteria}
    \label{fig:prisma}
\end{figure}

This SLR employed three different search processes to identify primary publications from the selected data sources outlined in Section \ref{section:datasource}: 1) automatic search on digital databases \cite{kitchenham2007}, 2) manual search on conference and journal proceedings \cite{kitchenham2007}, 3) bi-directional snowballing (identifying papers from the reference lists and the citations of a particular study) \cite{snowballing}. Then, the identified publications were evaluated using the selection criteria mentioned in Table \ref{selectioncriteria - table}. This evaluation resulted in a total of \textbf{39} suitable publications for this SLR, which are depicted in Table \ref{table:selectedstudies} along with their references. The publications are sorted alphabetically according to the first author' last name. The three search and selection strategies were initially conducted during May-July 2023. However, to incorporate the recent developments, we subsequently performed a second round of search and selection activities in November 2023. Figure \ref{fig:prisma} illustrates the search and selection strategy of this SLR. This figure also states the number of articles excluded and the reasons for exclusion.

\begin{table*}
  \caption{The list of the 39 publications selected for the SLR.}
  \label{table:selectedstudies}
  \begin{tabular}{llll}
    \toprule
    \multicolumn{1}{c}{Authors} & Reference & \multicolumn{1}{c}{Authors} & Reference\\
    \midrule
    Agrawal et.al (2021) & \cite{privacyworlds} & Kek\"{u}ll\"{u}o\u{g}lu and Acar (2023) & \cite{startuptocore} \\
Alhazmi and Arachchilage (2021) & \cite{allears} & Klymenko et.al (2022) & \cite{technicalmeasures} \\
Alkhatib et.al (2020) & \cite{notyet} & Li et.al (2018) & \cite{coconut} \\
Alomar and Egelman (2022) & \cite{darn} & Munilla et.al (2023) & \cite{lessonslearned} \\
Baldassarre et.al (2022) & \cite{pkbtool} & Peixoto et.al (2023) & \cite{brazilian} \\
Barbala et.al (2023) & \cite{agile} & Perera et.al (2020) & \cite{privacyawareiot} \\
Berk et.al (2023) & \cite{gamedevs} & Peretz et.al (2021) & \cite{climate} \\
Boenisch et.al (2021) & \cite{neverthought} & Ribak (2019) & \cite{culture} \\
Chen et.al (2018) & \cite{aswegrow} & Rocha et.al (2023) & \cite{lgpd} \\
Cobigo et.al (2020) & \cite{cognitivedisability} & Sanderson et.al (2023) & \cite{aiethicspractice} \\
Collier and Stewart (2022) & \cite{privacyworlds} & Sarathy et.al (2023) & \cite{dontlook} \\
Dias et.al (2020) & \cite{ictperception} & Senarath and Arachchilage (2018) & \cite{devcannotembed} \\
Dwork et.al (2019) & \cite{epsilon} & Senarath and Arachchilage (2018) & \cite{dmapproach} \\
Ekambaranathan et.al (2021) & \cite{moneymakes} & Senarath et.al (2019) & \cite{willtheyuse} \\
Greene and Shilton (2018) & \cite{platformprivacies} & Spiekermann (2018) & \cite{insidetheorg} \\
Hadar et.al (2018) & \cite{mindset} & Stahl et.al (2022) & \cite{ethicalissues} \\
Hargitai et.al (2018) & \cite{goingbeyond} & Tahaei et.al (2021) & \cite{privacychampions}\\
Horstmann et.al (2023) & \cite{thosethings} & Zhang et.al (2018) & \cite{towardsfl} \\
Ib\'{a}\~{n}ez and Olmeda (2022) & \cite{aiethics} & & \\
Iwaya et.al (2023) & \cite{inthewild} & & \\
    \bottomrule
  \end{tabular}
\end{table*}

\paragraph{Digital database search} An automatic search was performed on the six selected digital databases. The advanced search features of the digital databases and the search strings presented in Table \ref{stringsset} were used to create a refined \emph{search query}. In this search query, \textbf{quotation marks (" ")} were used to enclose multiple words together, ensuring that the search results included the exact match (e.g., "software engineer"). In addition, the \textbf{asterisk wildcard (*)} replaced any unknown characters in a string. For instance, the term \textbf{developer*} was used to retrieve studies that contain the words `developer' or `developers'. Further, the `AND' and `OR' Boolean operators were used to construct a more sophisticated and logical query. The final search query used to search the digital libraries is as follows:

\hfill \break
\noindent
\emph{
\scriptsize(developer* \textcolor{magenta}{OR} programmer* \textcolor{magenta}{OR} "software engineer" \textcolor{magenta}{OR} "software engineers" \textcolor{magenta}{OR} "software community" \textcolor{magenta}{OR} "software industry" \textcolor{magenta}{OR} "software system" \textcolor{magenta}{OR} "software systems" \textcolor{magenta}{OR} "software creator" \textcolor{magenta}{OR} "software creators" \textcolor{magenta}{OR} coder*) \textcolor{blue}{AND} (PETs \textcolor{magenta}{OR} "privacy enhancing technology" \textcolor{magenta}{OR} "privacy enhancing technologies" \textcolor{magenta}{OR} "privacy preserving technology" \textcolor{magenta}{OR} "privacy preserving technologies" \textcolor{magenta}{OR} "privacy enhancing technique" \textcolor{magenta}{OR} "privacy enhancing techniques" \textcolor{magenta}{OR} "privacy preserving technique" \textcolor{magenta}{OR} "privacy preserving techniques" \textcolor{magenta}{OR} "privacy protection technology" \textcolor{magenta}{OR} "privacy protection technologies" \textcolor{magenta}{OR} "privacy protection technique" \textcolor{magenta}{OR} "privacy protection techniques" \textcolor{magenta}{OR} "data protection technologies" \textcolor{magenta}{OR} "data protection techniques" \textcolor{magenta}{OR} anonymisation \textcolor{magenta}{OR} anonymization \textcolor{magenta}{OR} pseudonymisation \textcolor{magenta}{OR} pseudonymization \textcolor{magenta}{OR} "synthetic data" \textcolor{magenta}{OR} encryption \textcolor{magenta}{OR} "zero knowledge proof" \textcolor{magenta}{OR} "differential privacy" \textcolor{magenta}{OR} "multi party computing" \textcolor{magenta}{OR} "federated learning") \textcolor{blue}{AND} (privacy \textcolor{magenta}{OR} "data protection") \textcolor{blue}{AND} ("software development" \textcolor{magenta}{OR} "developing software" \textcolor{magenta}{OR} "developing applications" \textcolor{magenta}{OR} "application development" \textcolor{magenta}{OR} "system development" \textcolor{magenta}{OR} "developing systems" \textcolor{magenta}{OR} programming \textcolor{magenta}{OR} coding \textcolor{magenta}{OR} "creating software" \textcolor{magenta}{OR} "creating applications" \textcolor{magenta}{OR} "software creation" \textcolor{magenta}{OR} "application creation")}
\break
\hfill

The database search resulted in a total of 1,867 publications. We then screened the abstracts of those publications to exclude the ones that did not satisfy the selection criteria. If the abstracts were not detailed enough to make a decision, we screened the full text of the study. After screening the publications from digital databases, we selected 22 publications, as shown in Figure \ref{fig:prisma}.

\paragraph{Conference and journal proceedings search} We manually searched the selected conferences and journals mentioned in Table \ref{proceedings} to identify additional publications which might have been overlooked during the digital database search. The search resulted in 70 publications, comprising 42 conference papers and 28 journal articles. Subsequently, seven publications were selected based on the selection criteria.

\paragraph{Bi-directional snowballing} We supplemented the search process with a snowballing search to expand the article coverage \cite{snowballing}. The 22 publications selected from the digital databases and seven publications selected from proceedings were used as the initial articles of the snowballing process. Then, from the reference lists (i.e., backward snowballing) and citations (i.e., forward snowballing) to those articles, ten new studies were selected.
\medbreak

During the search and selection process, it was necessary to assess the reliability of the selection decisions to minimise the researcher bias. Therefore, Cohen's Kappa statistic was used to measure the level of agreement between the authors \cite{kappa}. This statistic resulted in 0.88, which is a nearly perfect agreement. The authors then discussed the discrepancies and took appropriate actions to resolve them. For example, there was a disagreement about including publications that did not explicitly mention the term `PETs' yet discussed developers' technical privacy behaviour. The authors discussed and collectively decided that those papers satisfied inclusion criteria (IC1) in Table \ref{selectioncriteria - table}. This decision was taken because those studies might indirectly shape how the software developers perceived and incorporated PETs in their software. For instance, software developers who lack knowledge of privacy regulations might not know that PETs can be used as a technical solution to adhere to those regulations.

\subsubsection{Data Extraction and Analysis}
The authors employed the reflexive thematic analysis proposed by Braun \& Clarke \cite{TA} to analyse the selected primary publications. Reflexive thematic analysis encourages researchers to engage with their own experience, knowledge, and perspective to delve deeper into the data and extract rich meaning behind them. This method particularly suits our study due to the diverse expertise of the authors. All authors have industrial experience as software developers, and some authors are well-established researchers in software privacy. The reflexivity offered by this dual perspective forms a strong foundation for our study, which aims to explore PETs in the context of software development. 

Before the analysis phase, the authors extracted data from the selected primary studies and recorded them in a tabular format which included the following fields. 

\begin{itemize}
\item General information: \emph{title, authors, author’s contact details, published date, published venue, reference, notes}
 \item Study Design: \emph{study type, methodology, details about the data set, follow up experiments}
 \item Data fields specific to the underline research questions
 \item Other data: \emph{limitations, future work, additional notes, our insights}
\end{itemize}

{\setlength{\parindent}{0pt} We examined the extracted data to familiarise ourselves with the content of the selected publications \cite{TA}. This familiarisation process allowed us to understand the key findings of the publications and build initial analytical thoughts on the common patterns and relationships in the findings. For example, after reading the extracted data, the authors discussed how the software development life cycle (SDLC) stages are related to the challenges in integrating PETs into software.}

After the familiarisation phase, the collected articles were open-coded (coding without a predefined coding scheme \cite{TA}) using NVivo\footnote{\url{https://lumivero.com/products/nvivo/}}, a qualitative data analysis software which provides interfaces and features to easily conduct and manage the processes in qualitative analysis. Both semantic codes (surface meaning) and latent codes (underpinning ideas behind the surface meaning) were generated during the coding process \cite{TA}. For example, the following excerpt semantically describes the users' inability to demand software privacy from the developers.

\begin{quote} 
\centering 
\emph{"Educate the market and end-users [about privacy], then developers will be forced to meet their requirements"} \cite{notyet} 
\end{quote}

\noindent However, we attributed this to a more latent code - "lack of perceived responsibility", as the developer was trying to pass the responsibility of protecting privacy to the end-users. In contrast, the following excerpt was attributed to a more straightforward (semantic) code - "unawareness of PETs".

\begin{quote} 
\centering 
\emph{"Differential privacy? No, I do not [know it]"} \cite{aiethics}
\end{quote}

Four iterative rounds of coding were conducted to extract as much meaning out of the publications. Then, the identified codes were organised under different themes based on their similarities and patterns. For example, we aggregated codes "evaluating performance" and "evaluating utility" under the theme "evaluation of PETs", which was related to explaining how developers assessed PETs. Subsequently, the themes were categorised under the defined research questions. The codes that did not fit the finalised themes were removed from the analysis process. Since using the concept of reflexivity led us to interpret the codes subjectively, applying inter-coder agreement to determine the final set of codes was not needed \cite{TA}. Instead, we collectively compared the codes and discussed their differences as the themes began to emerge.

\begin{figure}[hbtp]
   \centering
  \includegraphics[width=\linewidth]{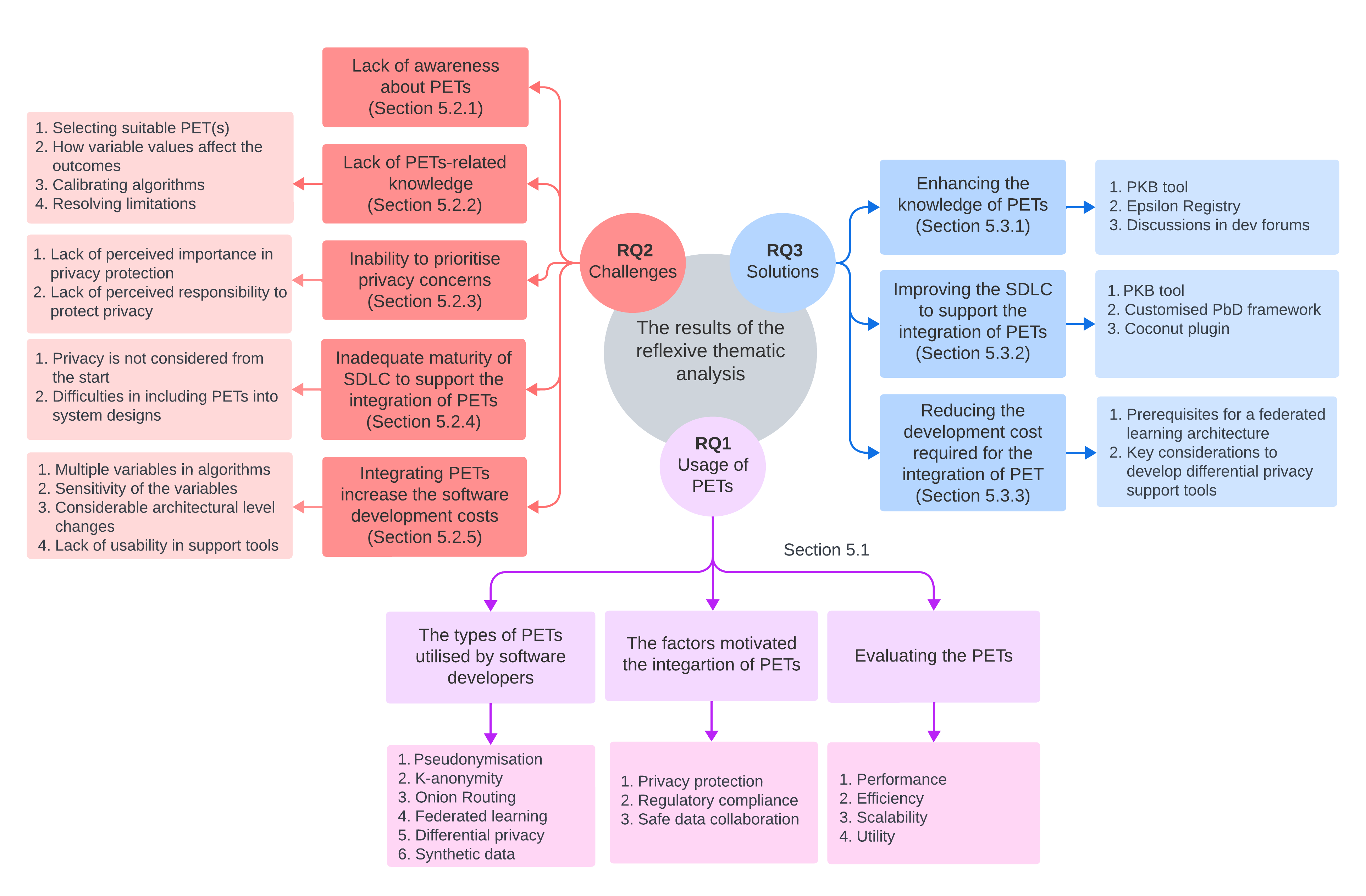}
   \caption{The overview of the themes identified from the reflexive thematic analysis of the Systematic Literature Review (SLR) and the main findings under each theme. The themes are categorised under respective research questions. The figure also mentions the sections that discuss each theme. RQ\# = Research Question}
   \label{fig:resultsfig}
\end{figure}

\section{Results} 
\label{results}

This section presents the results identified from the reflexive thematic analysis of the SLR. As shown in Figure \ref{fig:resultsfig}, we uncovered insightful findings into the types of PETs that developers have integrated into software and how they have integrated them (RQ1), the difficulties they encountered during the integration of PETs (RQ2), and the solutions introduced in the literature to address the identified difficulties (RQ3). 

\subsection{Integration of PETs into Software: Motivation, Application and Evaluation} \label{usage}

Our analysis revealed six types of PETs that developers have integrated into their software. These included long-established PETs, such as pseudonymisation \cite{goingbeyond}, k-anonymity \cite{goingbeyond}, and onion routing \cite{privacyworlds}. In addition, developers also integrated emerging PETs including, federated learning \cite{towardsfl}, differential privacy \cite{epsilon}, and synthetic data \cite{dontlook}. Software developers integrated the identified PETs into software motivated by three main reasons: the end-user privacy protection \cite{epsilon,privacyworlds,towardsfl}, regulatory compliance \cite{culture, petsgovernanddesign} and opportunity for secure data collaboration \cite{aiethicspractice, goingbeyond}. Further, they used evaluation criteria, including scalability \cite{towardsfl}, efficiency \cite{towardsfl}, and performance \cite{privacyworlds, towardsfl}, to assess PETs before selecting them or during their integration. The identified application scenarios demonstrated the privacy-preserving capabilities of PETs in different software domains and fields. These included healthcare \cite{aiethicspractice}, finance \cite{goingbeyond}, legal \cite{goingbeyond}, government services \cite{goingbeyond}, telecommunication \cite{towardsfl, privacyworlds} and machine learning (ML) \cite{aiethicspractice}. These results are summarised in Table \ref{tab:rq1-results} with references to the respective publications. In the following subsections, we report these findings in detail.

\begin{table*}
  \caption{The instances where Privacy Enhancing Technologies (PETs) were integrated into software by developers. The table depicts the PETs used in the reported instances, the empirical studies that reported the instances, the reasons for using the reported PETs, the related software domains related to the application instances, and the criteria used to evaluate the reported PETs. "-" was used when the studies did not provide specific details.}
  \label{tab:rq1-results}
  \footnotesize
  \begin{tabular}{p{1.4cm}clp{2.1cm}p{5.5cm}l}
    \toprule
    PET & Reference & Reason(s) & Domain/field & How was the PET used & \begin{tabular}[c]{@{}c@{}}Evaluation \\ criteria\end{tabular} \\
    
    \midrule
    
    \texttt Pseudonymisation & \cite{goingbeyond} & privacy protection & legal & They had an application to evaluate legal papers and automatically identify personal data items, such as names and addresses. The system then relabelled the specified data using pseudonyms. & \multicolumn{1}{c}{-} \\\\
    
    \texttt & \cite{goingbeyond} & privacy protection & government services  & The sensitive data was masked. The developers were reluctant to explain the specifics of the application scenario fearing potential privacy risks. & \multicolumn{1}{c}{-} \\
    
    \hline \\
    
    \texttt K-anonymity & \cite{goingbeyond} & \begin{tabular}[c]{@{}l@{}}compliance \\ collaboration\end{tabular} & finance & Developers did not mention the fine details of the application but stated they used suppression and generalisation. The authors claimed the developers used a version of k-anonymity.& \multicolumn{1}{c}{-} \\\\
    
    \texttt & \cite{lessonslearned} & compliance & \multicolumn{1}{l}{-} & Generalised individual numeric values to a numerical range & \multicolumn{1}{c}{-} \\\\
    
    \texttt & \cite{lessonslearned} & \begin{tabular}[c]{@{}l@{}}compliance\end{tabular} & \multicolumn{1}{l}{-} & Removed personally identifiable information (PII) from the data sets & \multicolumn{1}{c}{-} \\\\
    
    \texttt & \cite{aiethicspractice} & privacy protection & machine learning & Removed personally identifiable information (PII) from the data sets & \multicolumn{1}{c}{-} \\
    
    \hline \\
    
    \texttt Onion routing & \cite{privacyworlds} & privacy protection & telecommunication & The user's internet traffic was bounced through a relay of intermediate servers so that the Internet Service Provider cannot identify the origin and destination of the communication. & performance \\ 
    
    \hline \\
    
    \texttt Federated Learning & \cite{aiethicspractice} & \begin{tabular}[c]{@{}l@{}}privacy protection \\ collaboration\end{tabular} & healthcare & Different research institutes collaboratively analysed genomic data without sharing each other’s data into a centralised location. & \multicolumn{1}{c}{-} \\\\
    
    \texttt & \cite{towardsfl} & \begin{tabular}[c]{@{}l@{}}compliance\end{tabular} & telecommunication &  Ericsson was planning to deploy differential privacy so that they can learn from their customers data which were distributed across different countries.& \begin{tabular}[c]{@{}l@{}}efficiency \\ scalability\\ performance\end{tabular} \\

    \hline \\

    \texttt Differential Privacy & \cite{aiethicspractice}, \cite{epsilon} & privacy protection                                                          & \multicolumn{1}{l}{-} & The study did not provide details of the integration process & \multicolumn{1}{c}{-} \\

    \hline \\

    \texttt Synthetic data & \cite{aiethicspractice} & privacy protection & machine learning & Replaced the machine learning datasets that included sensitive data. & \multicolumn{1}{c}{-} \\
    
    \bottomrule
  \end{tabular}
\end{table*}

\subsubsection{Pseudonymisation} An empirical study by Hargitai et al. \cite{goingbeyond} revealed two instances where developers used pseudonymisation. In the first instance, a developer from the legal sector explained how they developed software to automatically identify personal data items such as names and addresses and then relabel them using pseudonyms \cite{goingbeyond}. They were mainly motivated by the need to protect the public's privacy, as legal documents contain highly sensitive information regarding them. In the second instance, an application developed for government services also used pseudonymisation to mask sensitive data in its datasets \cite{goingbeyond}. These datasets were obtained from municipalities and contained sensitive data collected from the public. Therefore, to protect the public's privacy, developers anonymised the datasets using pseudonymisation. Further, they collaborated with their organisation's business, legal, and security units, and the customers to decide which sensitive data items to mask.

\subsubsection{K-anonymity} Three publications mentioned how developers used k-anonymity in software \cite{goingbeyond, lessonslearned, aiethicspractice}. Each reported application scenario used suppression, generalisation, or both to achieve k-anonymity.

According to an interview study conducted by Hargitai et al. \cite{goingbeyond}, developers working in the finance domain implemented k-anonymity to anonymise their datasets, primarily to comply with privacy regulations \cite{goingbeyond}. Apart from this primary motivation, they were also influenced by the ability to safely collaborate data with external parties for research or business purposes. Such collaborations are otherwise challenging due to privacy concerns \cite{aiethicspractice}. However, during these interviews, the developers did not provide specific details about the integration process of PETs. They only mentioned using a combination of suppression and generalisation techniques without revealing further information. In addition, the developers also mentioned that multidisciplinary actors, including legal, customer service and IT support, collaborated to decide the data fields that must be anonymised \cite{goingbeyond}.

In addition, an interview study by Garrido et al. \cite{lessonslearned} revealed two instances where developers used k-anonymity. In both instances, k-anonymity was used to abide by privacy regulations when using end-user data for secondary purposes. According to regulations such as GDPR, collected personal data cannot be used beyond their original purpose without the explicit consent of the data owners (e.g., GDPR article 5: personal data processing \cite{gdprpseudo}). To comply with this data protection principle, developers anonymised the datasets by mapping the individual numeric data items to numeric ranges (generalisation) and removing PII, including names and social security numbers(suppression). In both cases, the link between the data and their owners was removed, making those data no longer identified as `personal data', thus refraining from breaking the privacy regulations \cite{lessonslearned}. 

Further, an interview study conducted by Sanderson et al. \cite{aiethicspractice}, which investigated the ethical practices of artificial intelligence researchers and software developers, also mentioned the usage of suppression. During this study, a developer mentioned using suppression to eliminate PII from datasets to protect the privacy of the data owners. However, further details about the application of suppression (e.g., data fields that the developers suppressed) were absent.

\subsubsection{Onion routing} Only one publication discussed how software developers used onion routing \cite{privacyworlds}. A case study conducted by Collier et al. \cite{privacyworlds} interviewed developers of the Tor browser to explore their privacy practices. Fearing how the Internet can threaten the privacy of its users, these developers invented the Tor browser using onion routing. In Tor, onion routing directs users' Internet traffic over a network of intermediary servers before reaching its destination. Due to this approach, the origin and destination of the communications are kept hidden from the Internet Service Provider (ISP). As a result, the users could browse the Internet anonymously, which protected their privacy \cite{privacyworlds}. 

During this empirical study, developers expressed that they defined a threat model \footnote{identify a set of risks and their mitigation discourses by analysing particular use cases, attacks, and defences \cite{privacyworlds}} before developing the Tor browser. Such threat modelling allowed the developers to make the best development decisions to improve the anonymity of the browsing experience. In addition, the study mentioned that developers used performance as a metric to evaluate the onion routing technique. For instance, during the development of the Tor browser, developers thought of including "padding traffic", i.e., fake traffic, to elevate the anonymity offered by the onion routing technique. However, they later decided to discard this idea as it would have created a significant performance overhead, slowing down the browsing experience of the users \cite{privacyworlds}.

\subsubsection{Federated Learning} Two empirical studies reported how developers used federated learning in software \cite{aiethicspractice,towardsfl}. Sanderson et al. \cite{aiethicspractice} reported a scenario in which federated learning was incorporated into a healthcare application. In this scenario, multiple health institutions wanted to collaboratively analyse genomics data, which holds a wealth of insights about humans. However, since genomics data are highly sensitive personal data, the institutions did not want to share their data. Hence, by using federated learning in the analysis process, developers protected the privacy of the data owners while allowing the institutions to gain insights from the genomics data \cite{aiethicspractice}.

In addition, a case study conducted with Ericsson's developers highlighted how they were planning to use federated learning to ensure regulatory compliance \cite{towardsfl}. Ericsson, a leader in the telecommunication field, could not transfer their worldwide customer data to a centralised location due to strict privacy regulations in some countries, which restricted moving data out of their jurisdiction. To address such regulations, the developers planned to utilise federated learning, enabling them to gain insights from decentralised customer data.

Further, developers of Ericsson evaluated federated learning \cite{towardsfl} using multiple criteria before incorporating it. They were concerned about smoothly managing resources (e.g., communication bandwidth) when remote clients grow, i.e., scalability assessment. In addition, this potential growth in the customer base has led them to assess the efficiency of federated learning. With the increase in remote clients, developers anticipated high communication congestion with the global machine learning model, leading them to consider communication efficiency. Moreover, they were also concerned about how the data heterogeneity in remote clients could impact the performance of the global model.

\subsubsection{Differential Privacy} None of the empirical studies included in our SLR provided comprehensive real-world scenarios for integrating differential privacy. Only two instances were found where developers simply acknowledged using differential privacy in their applications without providing additional details \cite{epsilon, aiethicspractice}. 

However, some empirical studies reported developers' opinions about the accuracy of differentially private data \cite{lessonslearned,dmapproach}. Developers stated that the usage of differential privacy depends on the specific software use cases \cite{lessonslearned,dmapproach}. For example, a financial application interested in calculating the yearly budget requires a higher accuracy, thus eliminating the use of differential privacy since it adds noise to the generated query results \cite{lessonslearned}. In addition, some developers commented on how the data quality can decide whether or not to use differential privacy. For example, sensor-collected data, such as GPS locations, naturally has a slight inaccuracy to them \cite{lessonslearned}. Therefore, using differential privacy on such data further increases the inaccuracy by introducing additional noise, thus making the data unusable \cite{lessonslearned}.

\subsubsection{Synthetic Data} The usage of synthetic data was mentioned in one publication \cite{aiethicspractice}. According to the publication, ML developers chose synthetic data over datasets containing sensitive information, responding to privacy concerns associated with them. However, it was not mentioned how developers generated the synthetic datasets to mimic the properties of the replaced datasets. 

In addition, two selected publications suggested using synthetic data for testing the configuration of other PETs \cite{dontlook,lessonslearned}. For example, after configuring differential privacy, developers can use synthetic data to assess whether the expected accuracy level is achieved in the data query results. This approach eliminates the necessity to test differential privacy on actual data, thus further reducing privacy risks \cite{dontlook,lessonslearned}.

\subsection{Challenges Faced by Developers During the Integration of PETs} \label{devchallenges}

Our analysis uncovered challenges that either refrained developers from embedding PETs into software or made them integrate PETs ineffectively where desired data protection guarantees were not achieved. Some challenges were caused by internal factors stemming from developers' lack of awareness \cite{aiethics,aiethicspractice,agile} and knowledge of PETs \cite{dmapproach, dontlook} and the inability to prioritise privacy concerns in software \cite{aswegrow,moneymakes,darn,mindset}. On the other hand, some challenges were beyond developers' control, which included the lack of maturity of the software development life cycle (SDLC) to accommodate PETs into software \cite{inthewild, notyet,startuptocore} and the increased software development cost due to the inner complexities of PETs \cite{goingbeyond, petsgovernanddesign, aiethicspractice}.

\subsubsection{Lack of awareness about PETs} 

To integrate PETs into software, developers must first be aware of the existence of such technologies. However, our analysis revealed that some developers were still unaware of PETs \cite{aiethics,aiethicspractice, agile, dmapproach}. One publication indicated how developers directly expressed their unawareness of PETs:"Differential privacy? No, I do not [know it]" \cite{aiethics}. Sometimes, developers resorted to traditional PETs such as k-anonymity or pseudonymisation, even if unsuitable for a given application scenario \cite{goingbeyond}. They failed to acknowledge the existence of other PETs that would be more suitable for their application scenarios than the traditional PETs \cite{goingbeyond}. This preference towards traditional PETs was also witnessed during empirical studies where the developers were asked to embed privacy in a given experimental application scenario \cite{dmapproach}.

In other cases, developers' unawareness of PETs was displayed through their incompetent decisions to overcome privacy-related roadblocks encountered during software development \cite{aiethicspractice, agile, aiethics}. For instance, developers sometimes restricted the collection and usage of data to overcome the regulatory pressure \cite{aiethicspractice, agile}. This seemingly cautious approach displayed the "better safe than sorry" mentality \cite{agile} of the developers. However, the root cause of such behaviour can be traced to developers' lack of awareness about the existing PETs and their privacy-enhancing capabilities \cite{dmapproach}.

\subsubsection{Lack of PETs-related knowledge} \label{lackknowledge}
Some developers were aware of PETs but lacked the proper knowledge to understand and integrate them effectively to address privacy concerns within their software projects \cite{dmapproach, dontlook}. Developers struggled to decide which PETs to use, when to apply PETs, and what privacy level they should aim to achieve when applying PETs in a given scenario \cite{dmapproach}. Responding to these uncertainties, developers sought external verification for their decisions, displaying their lack of confidence in integrating PETs into software \cite{dmapproach, dontlook}. 

Additionally, when incorporating PETs with an algorithmic background into software, developers struggled to understand how to set the parameters of the algorithms, how these parameters affect the entire application (e.g., impact on performance and utility) and what parameters must be changed to achieve the desired privacy levels \cite{dontlook, lessonslearned}. For instance, some developers knew the concept of "noise" in differential privacy and how it affected their software's data accuracy and privacy levels. However, when initialising the noise value, they anchored to the other existing projects or used the parameter settings stated in research studies without deciding whether those noise values would suit their software applications \cite{epsilon}. 

Moreover, developers lacked the knowledge to resolve any vulnerabilities of the PETs deployed in their applications \cite{goingbeyond, dmapproach}. One such vulnerability is the de-anonymisation risk in pseudonymised data \cite{goingbeyond}. For example, L.Sweeney \cite{deanon} showed how anonymised medical datasets can be de-anonymised by identifying matching quasi-identifiers in publicly available voting lists. Developers aware of this vulnerability in their applications resorted to immature workarounds instead of taking measures to mitigate the vulnerability \cite{goingbeyond}. These workarounds included limiting access to the pseudonymised data within the organisation and maintaining the secrecy of the technical implementations \cite{goingbeyond}.

Further, aggravating the lack of knowledge in PETs, there is an absence of educational approaches for developers to learn about PETs \cite{privacychampions,coconut, privacychampions,petsgovernanddesign, aiethics}. Developers voiced their difficulties in accessing updated information on privacy concepts, a challenge that naturally extends to the realm of PETs \cite{privacychampions}. In addition, educational resources, such as research papers and online articles, contained complex information, making it difficult for non-experts to understand the fine details of PETs \cite{coconut, privacychampions,petsgovernanddesign}. Moreover, developers did not receive sufficient training opportunities to enhance their PETs-related knowledge \cite{aiethics, climate, ethicalissues}. For example, developers expressed that their organisations focused more on non-privacy-related training, such as avoiding workplace harassment \cite{aiethics}. Even when organisations initiated privacy training, those sessions lacked the delivery of technical aspects of achieving privacy \cite{climate}.

\subsubsection{Inability to prioritise privacy concerns}

As PETs come under the broader concept of privacy, developers' ability to prioritise privacy concerns is required to understand the need for PETs \cite{privacyworlds}. However, according to our analysis, some developers failed to prioritise privacy in software because they believed that protecting privacy is neither important  \cite{aswegrow,privacyawareiot, notyet, thosethings} nor their responsibility \cite{aswegrow, darn, inthewild}. 

Empirical studies reported that some developers fail to perceive the importance of addressing privacy concerns in their software \cite{aswegrow,privacyawareiot, notyet}. Developers believed that limiting the excessive collection and analysis of data due to privacy concerns was a roadblock to achieving unexplored benefits of data \cite{aswegrow,moneymakes,darn,privacychampions,climate}. Therefore, developers had reservations about implementing privacy protection principles, like data minimisation \cite{aswegrow,privacyawareiot,dmapproach}. In addition, they also believed that ensuring privacy protection in software is unnecessary as the end-users and the organisations did not demand enhanced privacy levels in software \cite{insidetheorg, aiethics,brazilian,inthewild,mindset,climate, ictperception}. For example, some organisations directly advised developers to ignore privacy concerns \cite{insidetheorg,inthewild,mindset,climate}, while some indirectly displayed their indifference to privacy by neglecting to provide adequate privacy training \cite{aiethics} or by failing to emphasise the need for regulatory compliance \cite{ictperception}. As a result, developers attempted to deliver a working product while overlooking privacy concerns, assuming that end-users and organisations value functionality more than privacy \cite{notyet}.

Further, developers who participated in the empirical studies often believed safeguarding privacy was beyond their job description \cite{brazilian, mindset, lgpd}. Hence, they were reluctant to take responsibility for integrating privacy into the software they developed. They held different views on who should safeguard privacy: software end-users \cite{aswegrow, darn, howdeveloperstalk,notyet,petsgovernanddesign}, upper management in the organisation \cite{darn,privacychampions,inthewild,mindset,culture}, specialised privacy teams \cite{darn,allears,notyet}, or mobile platforms into which they released software (e.g., Android Play Store) \cite{darn,platformprivacies}. Developers believed that user-based privacy measures, such as privacy policies and consent forms, are sufficient to protect end-user privacy \cite{darn,technicalmeasures,inthewild, cognitivedisability, culture, climate}. In addition, some developers believed that upper management should make privacy decisions and direct privacy concerns to dedicated privacy experts other than developers \cite{darn,technicalmeasures,inthewild, cognitivedisability, culture, climate, technicalmeasures,allears,notyet}. Additionally, developers assumed that marketplaces provided correct and complete privacy definitions and guidelines, limiting their exploration of privacy concepts \cite{platformprivacies}. Moreover, they had marketplaces accountable for automatically detecting privacy vulnerabilities with their software, overlooking the proactive mitigation of privacy issues \cite{darn}. These beliefs and actions demonstrated how developers passed on the responsibility of protecting user privacy to other parties rather than accepting that they also have a fair share in protecting it \cite{mindset}.

\subsubsection{Inadequate maturity of the software development life cycle (SDLC) to support the integration of PETs}

The SDLC is the systematic methodology that guides the decisions and actions of developers through the various stages of software creation, from requirement gathering to developing and maintaining the software \cite{sdlc}. However, our analysis indicated that the current SDLC practices were not transformed to support integrating PETs into software \cite{inthewild, notyet,startuptocore, technicalmeasures}. 

The privacy specifications were ignored during the requirement gathering stage, making privacy an afterthought in software development \cite{darn,notyet, inthewild}. A study by Peixoto et.al \cite{brazilian} reported how developers directly expressed that privacy was not a concern of their projects - \emph{"Never...in any project I participated in here (company), we focused on that (privacy)"} \cite{brazilian}. This lack of planning made developers improvise the technical integration of privacy, leading to a struggle with inadequate resources (e.g., time and developers) required for the process \cite{inthewild, privacyawareiot, brazilian}. One of the main contributors to ignoring privacy as a requirement was that the development teams were unaware of the potential privacy risks of their software \cite{brazilian,inthewild,gamedevs, notyet}. This outcome occurred due to the lack of standard risk assessment methodologies, such as privacy impact assessments (PIAs), used during software development \cite{howdeveloperstalk, notyet, startuptocore, devcannotembed}. Due to this absence of risk identification, developers failed to understand the need for technical measures, such as PETs, to mitigate potential privacy risks \cite{privacychampions}.

In addition, despite being the leading roles in software development, developers were not actively engaged in privacy-related discussions \cite{darn, inthewild}. Instead, decisions regarding implementing technical privacy measures, such as PETs, often came from higher management of the organisations \cite{technicalmeasures}. In some cases, this left developers powerless to take action through technical solutions such as PETs, even when they discovered privacy vulnerabilities in their systems \cite{inthewild}. 

Further, developers displayed limited skills in generating system designs with privacy in mind \cite{dmapproach, privacyawareiot}. As system designs are the blueprints for the development phase, overlooking privacy considerations during this stage leads to ignoring privacy concerns during the development stage \cite{dmapproach}. Two empirical studies reported how developers struggled to include adequate technical privacy measures, like PETs, when asked to create software designs incorporating privacy \cite{dmapproach,privacyawareiot}. This inability demonstrated that creating privacy-based designs, particularly those incorporating PETs, is not a well-established step in the SDLCs that developers follow \cite{dmapproach,privacyawareiot}.

\subsubsection{Integrating PETs increases software development cost} \label{effort}
Developers often aim to reduce the time-to-market of software to gain a competitive advantage \cite{privacychampions}. However, developers saw integrating PETs into software as an obstacle to this goal because of the increased software development cost incurred from the inner complexities of PETs \cite{petsgovernanddesign, dontlook,lessonslearned}. 

PETs, particularly the emerging ones, such as differential privacy and homomorphic encryption, contain many variables that need fine-tuning to achieve the required privacy guarantees \cite{petsgovernanddesign, lessonslearned}. Consequently, the variable adjustment is time-consuming and requires careful consideration, as each variable could significantly impact the outcomes of PETs \cite{lessonslearned, petsgovernanddesign}. Developers sometimes ignored PETs due to this difficulty, which negatively impacted the data protection achieved through the software they developed. For instance, during an empirical study conducted to explore organisational strategies to anonymise data \cite{goingbeyond}, one developer mentioned how their team considered synthetic data but later decided against it because they needed to adjust many variables to generate properties of the real datasets. In addition, some publications reported that minor alterations to the variables resulted in drastic changes to the performance, utility and even the privacy assurance of the entire application \cite{goingbeyond, petsgovernanddesign, aiethicspractice}. Due to this high sensitivity, developers needed to be extra cautious when calibrating the variables of PETs, not only during their initialisation but also when updating the applications \cite{petsgovernanddesign}.

Developers also mentioned how the existing software architectures make the integration of PETs more resource-consuming \cite{towardsfl,lessonslearned, petsgovernanddesign}. For instance, in federated learning architecture, data heterogeneity across different remote clients affects the performance of their learning models. As a result, developers required substantial time to rectify data standards in remote clients \cite{towardsfl}. 

Tools like libraries \cite{sealcrypto,emptoolkit,googledp}, frameworks \cite{privc, cohort}, or visualisation aids \cite{psi,vip} exist to streamline the integration of PETs into software. However, developers who are not experts in PETs found it difficult to use these tools \cite{petsgovernanddesign, dontlook}. Developers highlighted issues with PET-related libraries, citing concerns about the ideal level of information abstraction \cite{dontlook,petsgovernanddesign}. Some libraries exposed excessively complex operations, making it challenging to create customised applications \cite{petsgovernanddesign}. In contrast, others lacked sufficient background information for informed decision-making, presenting difficulties for non-experts \cite{dontlook,petsgovernanddesign}. 

In addition, two interview-based studies found that some developers faced difficulties using tools developed to integrate differential privacy into software \cite{dontlook,lessonslearned}. These challenges stemmed from the tools not aligning with developers' current data analysis workflows \cite{dontlook,lessonslearned}. They lacked options to visualise the data query results and their accuracy levels after adding noise \cite{dontlook,lessonslearned}. Moreover, they excluded options to test differential privacy configurations using synthetic datasets \cite{dontlook,lessonslearned}. Further, these tools lacked sufficient export options for generating outputs in multiple file formats, such as comma-separated values (CSVs) and Jupyter Notebook files, which are commonly used for data analysis \cite{dontlook}.

\subsection{Proposed Solutions to Resolve the Identified Challenges} \label{solutions}
The literature suggested solutions to address only three challenges identified in Section \ref{devchallenges}: lack of PETs-related knowledge, inadequate maturity of SDLC, and reducing the development cost of integrating PETs into software. We also captured the limitations of these solutions to identify the areas for improvement. 

\subsubsection{Enhancing the knowledge of PETs}

The literature suggested a PETs-related knowledge base \cite{pkbtool}, knowledge-sharing mechanisms \cite{epsilon,howdeveloperstalk}, and privacy tutorials and frameworks with a technical touch \cite{lgpd,dontlook} to enhance the developers' knowledge about PETs.

The Privacy Knowledge Base (PKB) tool developed by Barletta et al. \cite{pkbtool} aimed to support developers' decisions during privacy-oriented software development. This tool allows developers to enter a software vulnerability they require to resolve along with the type of architecture of their software (e.g., client-server architecture). Then, the tool suggested suitable PETs to address the entered vulnerability. Using the PKB tool, developers can gradually learn how PETs can be helpful in data protection by mitigating software vulnerabilities. Therefore, we considered this tool as an approach to enhance the developers' knowledge of PETs, in addition to its original goal of supporting privacy decisions. However, the tool did not cover more advanced knowledge concepts of PETs, such as to integrate the selected PETs into software technically. Furthermore, even though the tool's usability and operability were evaluated, its effectiveness in educating developers is yet to be assessed \cite{pkbtool}.

In addition, two solutions emphasised learning about PETs through knowledge-sharing. Dwork et al. \cite{epsilon} proposed the "Epsilon Registry," a knowledge archive into which organisations who have already deployed differential privacy can enter the details of their integration scenarios. This registry allows developers to learn from the experiences and best practices of others who have already integrated differential privacy into software, such as configuring the noise value. However, organisations may have low incentives to disclose such knowledge voluntarily \cite{epsilon}. A similar behaviour was also seen in a study conducted by Hargitai et al. \cite{goingbeyond}, where developers were reluctant to disclose the details of their implementation of anonymisation techniques. In such instances, Dwork et al. \cite{epsilon} suggested using the coercive power of law to influence information disclosure. Another knowledge-sharing solution proposed by Li et al. \cite{howdeveloperstalk} suggested online developer forums to provide templates for privacy-related posts and encourage developers to exchange their privacy practices. The authors suggested that the forums establish guidelines that prompt developers to ask for better data practices from other developers. However, the identified knowledge-sharing solutions were conceptual and not evaluated through experimental studies \cite{howdeveloperstalk}.

Furthermore, some publications suggested readily available tutorials to enhance the knowledge of PETs \cite{dontlook,allears}. Sarathy et al. \cite{dontlook} discussed that the content of such tutorials should be simple, accompanied by code snippets, mathematical examples and visualisation mechanisms for better understanding. In addition, researchers advocated the design of privacy tutorials and frameworks to associate regulatory principles with relevant technical measures \cite{lgpd}. For instance, Rocha et al. \cite{lgpd} proposed a framework suggesting technical approaches to achieve principles in the Brazilian General Data Protection Law (LGPD) \footnote{Brazilian General Data Protection Law is the English translation of Lei Geral de Proteção de Dados Pessoais, LGPD, in Portuguese \cite{ictperception}}. For example, this framework suggested using anonymisation for the `data security' principle of LGPD.  However, the framework failed to provide the exact PETs to be employed (e.g., PETs to achieve anonymisation).

\subsubsection{Improving the SDLC to support the integration of PETs}
Three studies provided solutions to transform the SDLC to support the integration of PETs into software \cite{pkbtool, privacyawareiot, coconut}. These solutions focused on the design and development stages of the SDLC.

Our analysis stage identified two solutions that guide developers in creating privacy-oriented system designs focusing on PETs \cite{pkbtool,privacyawareiot}. The PKB tool described in Section \ref{lackknowledge} also suggested privacy patterns, which are practical design solutions to address privacy concerns. The PETs suggested by the tool are mentioned alongside the privacy patterns to convey how each pattern is achieved technically. Therefore, we categorised the PKB tool as an approach towards creating system designs by considering PETs. The second solution, presented by Perera et al. \cite{privacyawareiot}, included a set of privacy by design guidelines customised to address privacy concerns in five stages of the IoT data workflow (acquisition, preprocessing, processing and analysis, storage, and dissemination). For example, `hidden data routing' was a guideline to anonymise personal data during the data acquisition and dissemination stages. However, this framework only suggested the expected privacy-preserving functionality (e.g., hidden data routing) rather than mentioning the specific PETs to utilise (e.g., onion routing). 

Moving on to improving the development stage of the SDLC, we identified the Coconut - Android Studio plugin, which aimed to guide developers in resolving privacy risks during software development \cite{coconut}. It provided a way to document data handling practices through customised annotations. These annotations reported aspects such as the purpose, frequency, granularity, and visibility of a data-handling code snippet. The study did not explicitly state how PETs were supported by the tool, but we captured such instances through its usability experiment. For example, when a developer attempted to collect fine-grained location data, the plugin provided a real-time warning, indicating that coarse-grained location collection was sufficient. In this instance, the coarse-grained location collection is a location anonymisation technique achieved through generalisation \cite{locationtrunc}. However, the controlled nature of the usability experiment and the limited participation of only 18 developers limited the applicability of the findings to a larger population of developers \cite{coconut}.

\subsubsection{Reducing the development cost required for the integration of PETs} \label{reducingcost}

We identified three solutions proposed to reduce the software development cost of integrating PETs into software \cite{towardsfl, lessonslearned, petsgovernanddesign}. However, none of these three solutions were evaluated in the software development context.

Zhang et al. \cite{towardsfl} proposed five prerequisites to consider in a software architecture before designing a federated learning system. These included providing services for continuous real-time model learning, efficient resource utilisation for model training at remote clients, streamlined communication with central servers, performance evaluation mechanisms for machine learning models in remote clients, developing scalable architectures, and establishing secure connections between edge clients and the central server. By adhering to these criteria, developers can reduce misunderstandings and revisions during the integration of federated learning.

In addition, two solutions focused on enhancing developers' usage of PETs based support tools \cite{lessonslearned, petsgovernanddesign}. Garrido et al. \cite{lessonslearned} studied the developers' standard data analysis workflow and then defined ten key desiderata that differential privacy tools (e.g., libraries) should satisfy. Their effort was to align the functionality of such tools with the current data analysis workflow of developers. For example, the authors proposed that differential privacy tools should provide ways to visualise the data query results and the accuracy achieved. Moreover, Agrawal et al. \cite{petsgovernanddesign} suggested a two way approach to improver Moreover, Agrawal et al. suggested a two-way approach to improve the adoption of PETs related support tools. e the adoption of PETs related support tools. Since the existing libraries for PETs are complex and suited for expert usage, the authors suggested that experts in PETs first use the libraries and design components for common operations. Then, non-experts should integrate those components into systems.

\section{Discussion} \label{discussion}

This SLR analysed 39 empirical studies by following the guidelines of Kitchenham and Charters \cite{kitchenham2007} to understand how developers can be encouraged and guided to incorporate PETs into software. For this purpose, we first uncovered how developers currently using PETs (RQ1), such as pseudonymisation \cite{goingbeyond}, k-anonymity \cite{goingbeyond}, onion routing \cite{privacyworlds}, differential privacy \cite{epsilon}, federated learning \cite{towardsfl}, and synthetic data \cite{dontlook}. Then, we discovered the challenges they face while integrating PETs (RQ2), ranging from internal challenges like limited awareness \cite{aiethics}, insufficient knowledge \cite{dmapproach}, and inadequate emphasis on privacy \cite{aswegrow} to external obstacles such as a lack of SDLC support \cite{inthewild} and increased development costs \cite{petsgovernanddesign}. Finally, we identified solutions (RQ3) proposed to address the uncovered challenges in RQ2 \cite{pkbtool, coconut}, which had limitations around their comprehensiveness and generalisability. 

These results led to three implications. First, comparing developers' current usage of PETs with the commendable research conducted around PETs, it is questionable whether the broader developer community widely and satisfactorily uses PETs. Second, the challenges of integrating PETs express a need to identify directions to encourage the acceptance and usage of PETs within developers. Third, concrete solutions are required to address the identified challenges effectively, integrating seamlessly with developers' workflows and preferences for heightened usefulness.

In this section, we delve into these implications, suggesting future research avenues to identify research gaps. Further, we outline the limitations of the SLR and the attempts taken to reduce them.

\subsection{Research Efforts in PETs and Their Reflection on Software Developers} \label{notablegap}

From the results discussed in \ref{usage}, we observed some disparities between the theoretical advancements of PETs and their translation in developers' privacy practices. As presented in Table \ref{tab:rq1-results}, one such significant observation was that developers used a limited variety of PETs in a limited number of instances. This observation was unexpected given the wide range of PETs and extensive application scenarios discussed in PETs-related literature \cite{21,39, tippers,fate,exdra, controltracing}. Some of the PETs that were not investigated in the empirical studies are homomorphic encryption \cite{medco, videosharing, exdra, cohort}, secure multi-party computation \cite{fate,exdra,recordlinkage}, zero-knowledge proof \cite{controltracing}, and trusted execution environments \cite{prochlo, Ryon}, which are known to be highly applicable in the privacy-preserving software development context. Further, none of the empirical studies reported PETs-embedded mobile software applications. This oversight was surprising since the mobile application domain is a leading software domain \cite{aswegrow, howdeveloperstalk} in which the incorporation of PETs has already proven feasible \cite{videosharing, felicitas, mobilemental,mobhide}. 

Another recurring trend was the lack of detailed explanations of the steps developers followed when integrating PETs into software. The empirical studies did not report why and how developers selected the corresponding PET(s) for their applications, how they addressed the limitations of PETs, if any, and how they tested the configured PETs. Even when explaining the evaluation process of PETs, the publications only reported the evaluation criteria that developers considered and did not inquire into the quantitative or qualitative methods used to evaluate PETs against other system requirements (e.g., reduced response time to user requests \cite{medco}). It is also important to note that none of the publications mentioned how developers evaluated the `privacy levels' achieved through PETs, even though `privacy protection' was the most cited motivation to integrate PETs, as shown in Table \ref{tab:rq1-results}. 

This abstract information contrasts with the literature on PETs, where researchers offer more insightful descriptions of the approaches taken while integrating PETs into software \cite{medco,mobilemental,saphana,tippers,mobhide, exdra}. For example, the authors of MEDCO \cite{medco}, a system to explore distributed clinical data using homomorphic encryption and differential privacy, explained how they quantitatively evaluated the scalability and efficiency of the system by changing the data query sizes and dataset size. However, such details regarding the integration of PETs done by developers were absent in the selected publications. 

Furthermore, empirical studies made minimal effort to explore the existence and usage of the tooling support to simplify the integration of PETs. As discussed in Section \ref{devchallenges}, a few developers were aware of the limitations of such tooling support \cite{petsgovernanddesign}; however, they did not mention specifically what tools they were referring to or whether they used those tools when developing their applications. Therefore, it is unclear how the broader developer community uses the libraries \cite{sealcrypto,emptoolkit,googledp}, frameworks \cite{privc, cohort}, and interfaces \cite{psi,vip} discussed in the literature. In addition, to make PETs easily blend into the development workflow, scientists have integrated PETs-related modules into the commonly used development systems such as Apache SystemDS (an ML system designed to cover the end-to-end data science workflow) \cite{apache} and SAP HANA (a relational database management system) \cite{saphana}. However, these research efforts were also not reflected in the real-life integration scenarios mentioned in the selected studies.

The selected empirical studies may have overlooked other integration scenarios of PETs, contributing to the disparity between research efforts and their translation in developers' data protection practices. There exist other large-scale commercial applications that use PETs, such as Google GBoard \cite{gboard}, Apple and Google's Exposure Notification Privacy-preserving Analytics System (ENPA) \cite{enpa}, and MyHealthMyData (MHMD) Horizon project \footnote{\url{http://www.myhealthmydata.eu/}} for privacy-preserving medical data sharing. However, such applications were not covered by the selected empirical studies. On the other hand, developers are sometimes reluctant to disclose the details of PETs embedded software, fearing privacy attacks against their integration \cite{goingbeyond}. Either way, the observed lack of detailed explanations leaves an open question about the adoption of PETs by the broader developer community, thus impeding the chances to uncover more avenues to encourage developers' behaviour in integrating PETs into software.

Therefore, we encourage researchers to conduct more empirical studies to better understand how developers incorporate PETs into software. If developers are reluctant to disclose details about their prior experiences, the studies can provide hands-on exercises to gain real-time observation of their competency in working with PETs. It is also noteworthy to consider the diversity of the participants along different aspects, such as expertise level, age, gender, nationality or community, privacy knowledge levels, and the region in which they work. This diversity supports the higher generalisability in the study results. Another way to improve the generalisability of the results is to investigate developers outside North American and European regions, which are rarely considered in existing empirical studies \cite{inthewild,culture}. Focusing on such understudied geographical areas also leads to identifying the role of cultural influence in the developers' data protection practices at a global scale.

\subsection{Are Developers Ready to Accept and Use PETs in Software Development?}

The challenges outlined in Section \ref{devchallenges} extend beyond conventional technical barriers, reaching into the psychological aspects of developers. These multifaceted challenges raise uncertainty about whether developers can accept and incorporate PETs into their development workflows. To closely examine this uncertainty, we turn to the Theory of Planned Behaviour (TPB), an analytical framework that sheds light on individuals' intention to engage in a particular behaviour \cite{plannedbe}. According to TPB, the intention to conduct a behaviour is influenced by three factors: attitudes (favourable or unfavourable evaluation), subjective norms (perception of others' view towards the behaviour) and perceived behavioural control (perceived ease or difficulty in performing the behaviour) \cite{plannedbe}. The following subsections discuss how the identified challenges affect these three factors.

\subsubsection{Attitudes} People tend to develop a positive attitude toward a behaviour when it produces favourable outcomes \cite{plannedbe}. However, the challenges outlined in Section \ref{devchallenges} indicate that developers hardly perceive favourable attitudes towards integrating PETs. When developers lack awareness and knowledge about the existing PETs, they might not fully understand the benefits of PETs, such as the privacy-protecting abilities, mitigating the risk of violating privacy regulations and enhanced business opportunities (e.g., collaboration). This perception causes the developers to have a neutral view towards integrating PETs. Further, the absence of a systematic methodology to integrate PETs (e.g., SDLC) and the increased software development cost can evoke a sense of burden among developers regarding the integration process of PETs \cite{dmapproach}. Therefore, the challenging nature of incorporating PETs into software makes it difficult for developers to instil favourable attitudes towards it.

\subsubsection{Subjective Norms} The social background of developers has a critical hand in shaping the developers' privacy perspective \cite{culture, insidetheorg, darn, moneymakes}. For example, studies claim that developers who are parents in real life have a higher perceived responsibility to protect privacy when developing applications for children \cite{moneymakes}. Therefore, developers might perceive protecting privacy in different ways. On the other hand, organisations that value consumer privacy or comply with privacy regulations can make developers work towards common privacy goals regardless of the differences in their perceptions \cite{climate}. However, our findings in Section \ref{devchallenges} speak otherwise. When organisations explicitly instruct developers to disregard privacy aspects \cite{insidetheorg,inthewild,mindset,climate} or fail to support developers to cater privacy concerns, developers might perceive a lack of value in privacy protection within their organisations. This perception could negatively influence their privacy practices, including integrating PETs into software. Additionally, when privacy requirements are not formally defined for a software application, developers might not fully comprehend the demand customers place on privacy and the need for PETs to satisfy those demands. 

\subsubsection{Perceived behavioural control} According to Ajzen \cite{plannedbe}, the intention to engage in a behaviour is positively influenced by the belief that it can be performed easily. The adequate resources and opportunities facilitating the behaviour determine the level of ease in conducting it \cite{plannedbe}. According to the challenges discussed in Section \ref{devchallenges}, developers find it difficult to experience such ease when integrating PETs into software. Developers lacked even the fundamental knowledge of PETs, such as selecting suitable PET(s) for a given application scenario \cite{dmapproach, dontlook,epsilon, goingbeyond}, which made them struggle to make informed decisions when incorporating PETs into the software they develop. In addition, the primary studies also discussed how the software development cost would increase due to the inner complexities of PETs requiring ample resources in terms of time, computational ability and expert support to mitigate such challenges. However, such privilege is absent for every developer, as some might work in start-ups or small-scale companies where resources are limited \cite{privacyawareiot,inthewild}. Therefore, the challenging nature of integrating PETs into software can make developers perceive it as a difficult behaviour to perform.

By analysing the identified challenges about the TPB, it is clear that developers' attitudes, subjective norms and perceived behavioural control have lower power in influencing the intention to develop PETs-embedded software. Therefore, concrete solutions are required to prepare developers to accept and use PETs as part of their development workflow. We also urge researchers to develop customised technology acceptance models to understand how developers can be influenced to accept and use PETs in the software development context, thus facilitating their adoption within the broader developer community. Such models can be influenced by existing technology acceptance models such as UTAUT (Unified Theory of Acceptance and Use of Technology) \cite{utaut} and TAM (Technology Acceptance Model) \cite{tam}.

\subsection{Mitigating the Challenges}

The identified solutions in \ref{solutions} that answer RQ3 were not concrete enough to address the challenges encountered by developers when integrating PETs. They were not entirely dedicated to resolving the challenge they attempted to answer. For example, for a PETs-based learning approach to be useful, it should answer the exact knowledge gaps that developers have regarding PETs. However, we did not observe such determination in most identified solutions. On the other hand, the identified solutions were not evaluated with a considerably large and diverse set of developers. Therefore, the applicability of such solutions in the developer community is questionable. TThis absence of concrete approaches to resolving the PETs-centric challenges the developers face can further discourage them from integrating PETs into software. Thus, in this section, we will provide insights to improve the identified solutions in Section \ref{solutions} to better align them with software developers' preferences and workflows. Further, we will suggest guidelines to mitigate the unanswered challenges identified in Section \ref{devchallenges}, such as developers' lack of awareness of PETs and inability to prioritise privacy concerns.

\subsubsection{Enhancing the awareness of PETs}

Awareness of PETs is essential for developers to know that privacy can be achieved technically. Otherwise, as discussed in Section \ref{devchallenges}, developers will resort to immature privacy practices \cite{aiethicspractice, agile} or use only traditional PETs \cite{goingbeyond, dmapproach} to achieve data protection in software. As several seminal works have suggested, a sensible place to start enhancing \emph{privacy} awareness would be through education provided in computer science course modules \cite{awareness, gamedevs, devcannotembed}. We suggest extending this idea by including PETs-related content in those modules. In this way, a proactive mindset could be built in developers towards integrating PETs in their software during their formative years as novices. Moreover, competitions targeted at promoting PETs, such as the "PETs prize challenges"\footnote{\url{https://petsprizechallenges.com/}}, should be actively promoted among the developer community. This initiative not only fosters a culture of innovation but also serves as a platform to raise awareness of PETs, encouraging developers to explore and understand diverse PETs and their potential applications.

In addition, organisations' ability to enhance \emph{privacy} awareness of developers through training sessions, code reviews and discussions between colleagues are suggested in privacy-related publications \cite{climate, privacychampions}. However, none of those publications emphasised including technical aspects, such as PETs, in those initiatives. Therefore, software organisations can consider further improving such initiatives by including PETs. In this way, the organisations can convince the technical feasibility of privacy using PETs, thus making the concept of PETs closer to developers.

Further, empirical studies sparingly analysed developers' awareness of different types of PETs and their capabilities. Thus, future research can provide a comprehensive understanding of developers' level of awareness in PETs, thereby promoting strategies to improve the awareness if needed. 

\subsubsection{Enhancing the knowledge of PETs}

The lack of knowledge in PETs is a fundamental yet significant challenge the developers face \cite{allears,devcannotembed}. The absence of theoretical and practical knowledge of PETs negatively impacts the developers' confidence in their ability, i.e., self-efficacy \cite{efficacy}, to effectively integrate PETs into software. As self-efficacy is a factor affecting the motivation to achieve desired goals \cite{seriousgame}, we argue that low self-efficacy arising from the absence of knowledge of PETs demotivates the privacy-preserving development behaviour of software developers, leading them not to integrate PETs into software or integrate them ineffectively \cite{goingbeyond, coconut, lessonslearned}. On the other hand, the researchers' effort in introducing new tools or methodologies to support the integration of PETs will be less useful since the developers fail to effectively utilise them without understanding PETs \cite{epsilon}. Thus, it is crucial to educate developers in a way that increases their self-efficacy to integrate PETs and their ability to utilise existing PETs-related support tools.

However, despite the importance of educating developers on PETs, none of the proposed educational approaches in literature were deemed. They should address the exact knowledge requirements of developers, such as selecting suitable PETs for software, calibrating PETs, evaluating the integration decisions, and rectifying the limitations found in PETs. Moreover, the educational approaches should equally focus on providing theoretical and practical knowledge about PETs. For this, they can design the teaching content into different levels according to Bloom's taxonomy \cite{bloom}, starting from lower learning levels, such as understanding concepts, to higher learning levels, such as applying the learnt concepts. In addition, they should consider the preferences that developers seek from educational resources, such as context-based learning \cite{gamedevs}, quick solutions (e.g., code snippets) \cite{coconut} and simple yet informative content delivery \cite{privacyawareiot}.

In addition, as a fundamental step towards enhancing the knowledge of PETs, we suggest educators incorporate PETs into ethics-related modules taught to computer science students (i.e., novice developers). Shilton et al. \cite{roleplaying} introduced a game-based learning tool to help computer science students understand how to achieve privacy in software applications by following privacy by design guidelines \cite{PBDandSE}. The authors experimentally showed that the tool influenced the students to learn more about privacy concepts during technical lessons. Therefore, further research can extend such interventions to include content on PETs, a part of the privacy domain.

Further, developers can be educated about PETs by providing adequate training sessions in software organisations \cite{devcannotembed}. For such training sessions to be useful, the delivered knowledge should be actively adapted to the organisational practices. Moreover, novel interventions, such as the previously discussed game-based approaches, can also be used in training sessions to provide developers with hands-on experience in utilising PETs. If such tools were made openly accessible, resource-limited organisations (e.g., startups) could also utilise them to educate their developers on PETs flexibly without needing extra time and money to organise special training sessions.

\subsubsection{Towards prioritising privacy concerns}

As identified in Section \ref{solutions}, developers failed to see the need for PETs as they did not prioritise privacy concerns during software development. This ignorance calls for organisations and regulatory bodies to develop strategies emphasising the significance of protecting privacy and how PETs can achieve privacy. 

Organisational privacy standards have the power to influence developers' privacy practices \cite{climate,ictperception,inthewild}. Developers usually follow organisational privacy policies to verify their privacy practices \cite{ictperception,inthewild}. Therefore, organisations should properly establish and communicate privacy policies with the developers. These policies can further include PETs, where personal data needs to be processed. Moreover, by maintaining systematic methodologies to integrate PETs (e.g., SDLC), organisations can convince developers that protecting privacy is part of their daily technical work. This approach also holds developers responsible for upholding software privacy standards throughout software development. In the future, more research can evaluate how organisations play a role in instigating the importance and responsibility of protecting privacy, motivating developers to use PETs to protect their consumers' data.

In addition, privacy regulations (e.g., GDPR) can convince the importance of protecting privacy to software developers. However, such regulations seem distant for developers as non-technical people, such as regulators, developed them \cite{thosethings}. Hence, privacy regulations can be presented with a technical touch to minimise this gap. For instance, they can include the types of PETs, their code examples and PETs-related tools that developers can use to achieve data protection. In this way, developers can identify the importance of addressing privacy concerns and the technical feasibility of privacy facilitated by PETs.

\subsubsection{Improving the SDLC to support the integration of PETs}

SDLC is essential in providing structured guidance to developers throughout the entire software development process, from requirement gathering to software deployment and maintenance \cite{sdlc}. Therefore, to successfully integrate PETs into software, every stage in the SDLC must actively support the incorporation of PETs. However, drawing from our results presented in Sections \ref{devchallenges} and \ref{solutions}, we noticed that past empirical studies have not attempted to elicit insights into how the integration of PETs is related to the testing, deployment, and maintenance stages of the SDLC. Therefore, future research can focus on the challenges and improvements along these stages, considering integrating PETs into software.

Privacy requirement gathering is the first step in the privacy-preserving software development process \cite{inthewild, devcannotembed}. However, none of the empirical studies provided well-structured approaches to elicit privacy requirements, which is required to identify the need for PETs. Highlighting the importance of prioritising privacy from the outset of software development, Iwaya et al. \cite{inthewild} proposed considering privacy as a non-negotiable feature. Additionally, Senarath and Arachchilage \cite{devcannotembed} proposed defining privacy requirements alongside the PETs needed to achieve them. However, these were only conceptual suggestions, not solutions with clear implementation steps. 

However, real-world development environments complicate privacy requirement gathering, requiring more than conceptual guidance. The collaborative contribution of different stakeholders (e.g., customers, investors, and developers) introduces challenges such as communication difficulties stemming from knowledge, terminology, and privacy expectation differences between the stakeholders \cite{ privacyimpact, aiethicspractice}. Further, privacy requirements extend beyond project initiation, requiring reassessment and adaptation throughout the SDLC \cite{privacyawareiot, privacyimpact}. Thus, future research should introduce concrete privacy requirement-gathering approaches that cater to such collaborative efforts and end-to-end management during software development.

Further, our analysis revealed a lack of explicit emphasis on documentation during the integration of PETs. The significance of documentation became apparent through examples like the 'Epsilon Registry' \cite{epsilon} and the Coconut plugin tool \cite{coconut}. The `Epsilon Registry' allowed for recording detailed decisions in successful differential privacy integration, providing valuable insights for other developers struggling to make integration decisions around differential privacy \cite{epsilon}. Similarly, the Coconut plugin employed annotations to document data handling behaviours, enhancing the transparency of the privacy-preserving coding practices \cite{coconut}. Therefore, we advocate that developers maintain robust documentation practices while integrating PETs. Further, researchers can explore the aspect of documentation more when developing approaches for a privacy-preserving SDLC.

Moreover, we noted a lack of SDLC-related solutions to support the integration of PETs in legacy systems. Developers working on legacy software must consider several other critical aspects when incorporating new components into their existing software, making the integration of PETs into them not straightforward. These include assessing the existing architectural limitations, identifying potential impacts on existing functionalities, determining resource allocation for the implementation, and addressing any interoperability challenges \cite{towardsfl}. Further, legacy software might need significant changes to incorporate PETs, which can increase software development costs \cite{towardsfl}. Such complexities can pressure developers not to integrate PETs into legacy software. Alternatively, developers may opt for PETs that impose minimal tension on the existing software architecture, even if such choices may not suit their applications. Therefore, future studies can reflect more on the unique challenges developers face when integrating PETs into legacy software.

Further, future studies can invest in developing an SDLC framework or a tool to support end-to-end privacy, of which PETs would naturally be a part. Microsoft Security Development Lifecycle \footnote{\url{https://www.microsoft.com/en-us/securityengineering/sdl}} is an industrial-level SDLC frameworks that guides security integration. Such interventions can be leveraged to develop a similar SDLC framework for privacy, also focusing on PETs. These research efforts should also encompass often-overlooked stages such as testing, deployment, and maintenance. In this direction, we suggest researchers investigate how different SDLC methodologies, such as Agile, Waterfall or V model \cite{sdlctypes}, can be adapted to incorporate PETs effectively. 

\subsubsection{Reducing the development costs}

Our findings also indicate that PETs researchers focus on achieving performance of PETs than their practicality in software development context \cite{technicalmeasures}. This is a probable cause for developers to struggle with the increased software development cost when integrating PETs. Drawing from the results presented in Section \ref{reducingcost}, an initial step towards enhancing the practicality of PETs is to gain insights into developers' existing development workflow. These insights can assist researchers in tailoring PETs and their supporting tools to align seamlessly with software development, reducing friction and enhancing adoption by developers. 

A practical example of this approach can be observed where functionalities required for deploying federated learning were integrated into Apache SystemDS, a machine learning system designed to cover the end-to-end data science workflow \cite{apache}. In this instance, embedding federated learning capabilities within an existing and widely used support system causes minimal disruptions to the ML workflow of developers. Additionally, using an existing system to support the integration of PETs eliminates the time and effort required to familiarise with a new federated learning support tool. Similarly, we encourage researchers to study the feasibility of incorporating the valuable findings of PETs in widely used development environments (e.g., Android Studio \footnote{\url{https://developer.android.com/studio}}), frameworks (e.g., AngularJS \footnote{\url{https://angularjs.org/}}), or database systems (e.g., Oracle \footnote{\url{https://www.oracle.com/}}) in the software industry. This way, developers can quickly adopt and integrate PETs into their existing workflows.

\subsection{Limitations}
This SLR consists of a set of limitations that can affect the validity of the study findings. In this section, we discuss these limitations under the validity classification proposed by Runeson and H\"{o}st \cite{threatstovalidity}, including construct validity, internal validity, external validity and reliability. Further, we explain the measures taken to minimise the identified limitations.

\subsubsection{Construct Validity} Limitations to construct validity can arise when the methods employed in a study do not accurately align with what the researchers intended to investigate \cite{threatstovalidity}. First, our research questions might not comprehensively explain why developers struggle to use PETs for privacy-preserving software development. We used the PICOC framework to mitigate this limitation \cite{slrsocial, kitchenham2007}, which guided us in identifying keywords to define well-structured research questions reflecting the study's scope. Second, the search strings we defined to identify publications might not be suitable to retrieve publications that answer the research questions. We used the main keywords derived from the PICOC framework as the primary search strings, to minimise this limitation. As these keywords are also components of the defined research questions, they can identify publications related to the defined research questions.

\subsubsection{Internal Validity} Internal validity is the extent to which the design and conduct of the study are likely to prevent systematic errors, such as missing relevant publications during search and selection \cite{kitchenham2007}. The search strings we defined might not retrieve all the relevant publications that answer our defined questions. To avoid this limitation, we first defined alternative search strings based on synonyms, abbreviations and spelling variations of the main search keywords we derived from the PICOC framework. Second, we conducted a rigorous search for publications in multiple data sources, including six digital libraries, 13 conferences, and 11 journals, so the relevant publications are less likely to be overlooked. In addition, we further expanded the coverage of the publications by manually searching the references and citations of the selected papers (snowballing method) \cite{snowballing}.

\subsubsection{External Validity} External validity refers to the extent to which study findings can be generalised \cite{threatstovalidity}. The generalisability of our findings might be limited due to the low number of selected empirical studies. Therefore, there is a need to conduct more research into PETs with a focus on software developers to validate our results. Nevertheless, our study will be a foundation for future investigations in this evolving and critical intersection of software development and PET.

\subsubsection{Reliability} Reliability refers to the extent to which the SLR depends on the specific researchers \cite{threatstovalidity}. Due to the background and expertise of the authors, this SLR could be biased towards specific researchers when selecting publications for analysis. To reduce this bias, the steps in the SLR methodology, including defining the research questions, creating the search strings, and defining the selection criteria and data sources, were all planned and verified among the author group before conducting the SLR \cite{kitchenham2007}. Further, the empirical studies selected for the SLR were validated using the inter-rater agreement score calculated between the authors using Cohen's Kappa measurement \cite{kappa}. Finally, during the reflexive thematic analysis, all authors used reflexivity to construct the codes and themes collaboratively to synthesise the results.

\section{Conclusion} \label{conclusion}
PETs are technical measures that offer data protection by design and default \cite{12, PETsICO}. Therefore, by embedding PETs into software applications, data breaches can be minimised, protecting the privacy of their end-users. Further, by improving the data protection in software through PETs, software organisations can manage regulatory pressure and enhance end-user trust \cite{devcannotembed,willtheyuse}. However, to benefit from the data protection capabilities of PETs, software developers must actively and correctly embed PETs into software. Therefore, to identify approaches to encourage such behaviour in developers, we conducted an SLR synthesising the knowledge regarding the developers' current usage of PETs, challenges they encounter during the integration of PETs and solutions proposed to address the identified challenges along with their limitations. Our SLR reported instances where developers incorporated six different PETs into software \cite{goingbeyond,aiethicspractice}. However, due to the absence of fine details in those scenarios, it is still unclear how the broader developer community adopts PETs. Moreover, developers encountered multiple challenges when integrating PETs into software, including personal challenges (e.g., limited awareness of PETs \cite{aiethics}) and external challenges (e.g., the increased development cost \cite{inthewild}). The identified challenges hint that the developers are still not ready to accept and use PETs in software. Additionally, existing solutions that address the identified challenges are either not entirely dedicated to addressing them or lack generalisability \cite{pkbtool,coconut,lessonslearned}. This absence of concrete solutions further aggravates the PETs-related challenges faced by developers. Based on the SLR findings and their implications, the adoption of PETs by the wider developer community may not be in the near future. Therefore, while dedicated efforts towards improving the capabilities of PETs are essential, it is equally important to understand how to improve developers' privacy-preserving development behaviour by taking PETs on board. We believe our study inspires this cause, encouraging researchers to investigate more into developer aspects in incorporating PETs into software, educators and practitioners to develop PETs-based learning interventions for developers, and organisations to identify and reduce the barriers integrating PETs within the development environment.

\bibliographystyle{ACM-Reference-Format}
\bibliography{sample-base}

\appendix

\section{Search Strings identified using the PICOC framework} 

\begin{table}[H]
\caption{The search strings used to identify the publications relevant to the research questions of the SLR. The PICOC framework was used to guide the creation of the search strings. Note that the * is used to represents any number of characters (e.g., engineer* can be either engineer of engineers)}
\label{stringsset}
\begin{tabular}{lp{10cm}}

PICOC element & Search strings                                                                                                                                                                                                                                                                                                                                                                                                                          \\ \hline
Population    & developer, programmer, software engineer*,  software community, software industry, software system*, software creator*, coder                                                                                                                     \\
              &                                                                                                                                           \\
Intervention  & PETs, privacy enhancing technolog*, privacy preserving technolog*, privacy enhancing technique*, privacy preserving technique*, privacy protection technolog*, privacy protection technique*, data protection technolog*, data protection technique*, anonymisation, anonymization, pseudonymisation, pseudonymization, synthetic data, encryption, zero knowledge proof, differential privacy, multi-party comput*, federated learning \\
              &                                                                                                                                                                                                                     \\
Outcomes      & data protection, privacy                                                                                                                                     \\
              &                                                                                                                                              \\
Context       & software development, developing software, developing applications, application development, system development, developing systems, programming, coding, creating software, creating applications, software creation, application creation                                                                                                                                                                                             \\ \hline
\end{tabular}
\end{table}

\end{document}